\documentclass[journal,10pt,onecolumn,draftclsnofoot,]{IEEEtran}
\IEEEoverridecommandlockouts
\usepackage[boxruled]{algorithm2e}
\usepackage{enumerate}
\usepackage{multirow,array}
\usepackage{dirtytalk}
\usepackage{cite}
\usepackage{graphicx}
\usepackage{psfrag}
\usepackage{caption}
\usepackage{subcaption}
\usepackage{url}
\usepackage{amsmath}
\usepackage{amsthm}
\usepackage{kbordermatrix}
\usepackage{array}
\usepackage{algorithm2e}
\usepackage{amssymb}
\usepackage{amsfonts}
\usepackage{float}
\usepackage{textcomp}
\usepackage{xcolor}
\usepackage{tabu}
\usepackage{array}
\usepackage{enumitem}
\newcolumntype{P}[1]{>{\centering\arraybackslash}p{#1}}
\newcolumntype{M}[1]{>{\centering\arraybackslash}m{#1}}

\def\BibTeX{{\rm B\kern-.05em{\sc i\kern-.025em b}\kern-.08em
    T\kern-.1667em\lower.7ex\hbox{E}\kern-.125emX}}
\usepackage{float}

\renewcommand{\boxed}[1]{\text{\fboxsep=.2em\fbox{\m@th$\displaystyle#1$}}}

\newcommand{\fq}{{\mathbb F}_q}
\newcommand{\gbinom}[2]{\boldsymbol{\bigl\langle}#1,#2\boldsymbol{\bigr\rangle}}

\newtheorem*{conjecture*}{Conjecture}
\newtheorem{lemma}{Lemma}

\newtheorem{corollary}{Corollary}
\newtheorem{remark}{Remark}
\newtheorem{definition}{Definition}
\newtheorem{theorem}{Theorem}
\newtheorem{example}{Example}

\title{Cache-Aided Communication Schemes via Combinatorial Designs and their $q$-analogs} 
\begin{document}

\author{

  \IEEEauthorblockN{Shailja Agrawal, K V Sushena Sree, Prasad Krishnan, Abhinav Vaishya, Srikar Kale}\\
  \IEEEauthorblockA{
                    International Institute of Information Technology, Hyderabad\\ 
                    Email: \{shailja.agrawal@research. , sushena.sree@research. , prasad.krishnan@, abhinav.vaishya@research. , srikar.kale@research. \}iiit.ac.in}

\vspace{-0.2cm}
}

\maketitle
\begin{abstract}
We consider the standard broadcast setup with a single server broadcasting information to a number of clients, each of which contains local storage (called \textit{cache}) of some size, which can store some parts of the available files at the server. The centralized coded caching framework, consists of a caching phase and a delivery phase, both of which are carefully designed in order to use the cache and the channel together optimally. In prior literature, various combinatorial structures have been used to construct coded caching schemes. One of the chief drawbacks of many of these existing constructions is  the large subpacketization level, which denotes the number of times a file should be split for the schemes to provide coding gain. In this work, using a new binary matrix model, we present several novel constructions for coded caching based on the various types of combinatorial designs and their $q$-analogs, which are also called subspace designs. While most of the schemes constructed in this work (based on existing designs) have a high cache requirement, they provide a rate that is either constant or decreasing, and moreover  require competitively small levels of subpacketization, which is an extremely important feature in practical applications of coded caching. We also apply our constructions to the distributed computing framework of MapReduce, which consists of three phases, the Map phase, the Shuffle phase and the Reduce phase. Using our binary matrix framework, we present a new simple generic coded data shuffling scheme. Employing our designs-based constructions in conjunction with this new shuffling scheme, we obtain new coded computing schemes which have low file complexity,  with marginally higher communication load compared to the optimal scheme for equivalent parameters. We show that our schemes can neatly extend to the scenario with full and partial stragglers also. 
\end{abstract}

\let\thefootnote\relax\footnotetext
{
Parts of this work appeared in the proceedings of the 2019 IEEE International Symposium on Information Theory (ISIT), Paris, France \cite{CDesigns_Shailja_ISIT2019} and at the 2020 IEEE International Symposium on Information Theory (ISIT), Los Angeles, CA, USA, \cite{Shailja_ISIT2020}.
} 

\section{Introduction}
Multi-terminal broadcast communication under the presence of side-information at the clients is a canonical setting for many modern communication problems, including satellite communication, 
content-delivery networks,  
and distributed computing. 
For the broadcast setting where each of the clients have local storage (called \textit{cache}), a joint caching-and-delivery design was proposed under the title of \textit{coded caching} in \cite{MaN}.  The coded caching scenario as in \cite{MaN}, consists of $K$ clients, each possessing some local storage (its \textit{cache}), connected to a single server through an error free broadcast link. The server contains a library of $N$ files of equal size, while each client cache can store content up to $M$ files (for some $M\leq N$). The coded caching framework in \cite{MaN} operates in two phases: the placement phase (in which the caches are populated) and the delivery phase. In the delivery phase (during peak hours), the demands of the users pop up. In the coded caching paradigm of \cite{MaN}, the server broadcasts \textit{coded} transmissions such that the demands of all the users are satisfied. As in \cite{MaN}, the rate $R$ of the coded caching scheme is defined as the ratio of the number of bits transmitted by the server to the size of each file. Designing a good coded caching scheme amounts to jointly designing the caching/prefetching phase, as well as a delivery scheme which can effectively make use of the coding opportunities resulting due to the cached subfiles at the clients, so that the rate of the scheme is minimized. 


For this coded caching system, the authors of \cite{MaN} presented a coded caching scheme which involved careful placement of subfiles of the files and an appropriately designed delivery scheme. The rate of the scheme in \cite{MaN} scheme was shown to be $R=\frac{K\left(1-\frac{M}{N}\right)}{1+\frac{MK}{N}}$, which has a  gain $1+\frac{MK}{N}=\Theta(K)$ gain over the uncoded delivery rate $\big($for constant $\frac{M}{N} \big)$. The rate achieved by Ali-Niesen scheme \cite{MaN} was shown to be optimal for a given cache size $M$ in \cite{WTP}, under the assumption of uncoded cache placement and $N\geq K$.

Interestingly, the paradigm set by \cite{MaN} was extended to a number of settings of cache-aided communications involving multiple terminals such as device-to-device communication (D2D) networks \cite{D2D}, distributed computing \cite{mapreduce}, and interference management in wireless interference channels \cite{interferencemanagement}.  One of the important settings where the coded caching paradigm has played a major role is in the popular distributed computing framework of MapReduce \cite{mapreduce}. In the MapReduce framework, a large data file is partitioned into smaller parts, and these parts are then assigned to different servers for processing in a distributed fashion. There are two main phases in MapReduce: \textit{map} and \textit{reduce}, and a third \textit{data shuffling} phase connects the two. In the map phase, each of the data parts is processed by one or more servers to generate some intermediate values (IVAs) using map functions. In the next step, servers exchange these IVAs (called \textit{data shuffling}) so that the final outputs can be calculated in a distributed fashion across the server using the {reduce functions}. As observed in \cite{33per} and \cite{70per}, data shuffling is a significant phase in determining the performance of the original MapReduce framework, which passes the IVAs in an uncoded manner during the shuffling phase.  In \cite{CMR}, it was shown that it is possible to code the IVAs together before the shuffling process by exploiting the fact that $r$ distinct carefully chosen nodes are mapping the same subfiles (and hence have the same IVAs). This leads to great savings in the communication load. The parameter $r$ is known as the \textit{computation load}, which indicates the price to pay for reducing the communication load. This new framework with a coded shuffling phase, is known as Coded MapReduce. This model was further studied in \cite{RDC}, where it was shown that the communication load achieved by the Coded MapReduce scheme of \cite{RDC} is optimal. In \cite{CommunicationComputationAlternativetradeoff,StorageComputationISWCS,QYDC}, the model proposed in \cite{CMR,RDC} was further extended to consider coded MapReduce schemes in which the nodes need not compute IVAs of all the stored subfiles for completing their reduce tasks. Tradeoffs between storage, computation, and communication loads were derived in \cite{StorageComputationISWCS,QYDC}, and an optimal scheme which meets this tradeoff was also presented.  Coded distributed computing schemes in the presence of \textit{stragglers} in the computation process were studied in \cite{Speedingup,UnifiedCodingFramework} for the case of computing functions which are linear. Stragglers are nodes which are either slow or completely unable to complete their map tasks. Subsequently, the works \cite{QYS,vinayakPDA}, extended the coded MapReduce model of \cite{RDC} to arbitrary function computation  in the presence of full and partial stragglers.

Though the Ali-Niesen schemes in coded caching \cite{MaN} and the Coded MapReduce scheme \cite{RDC} are shown to be optimal, the required \textit{subpacketization} level, which indicates the number of parts into which each file must be split into for enabling coded transmissions, increases exponentially in $K$ (for constant cache-fraction). In the design of the coded caching scheme in \cite{MaN}, each file must be split in this scheme into $\binom{K}{MK/N}$ subfiles, which can grow exponentially in the number of clients $K$ (for constant $M/N$) as $K$ becomes large. A similar issue arises in the coded distributed computing scenario also \cite{RDC}. This is a major setback for the scheme's practical implementation (see for instance, Section I in \cite{Suthan_etal_Subexp_ProjGeo} for a discussion on these issues). A number of research works have been introduced to address this problem in a variety of ways, including user-grouping based methods \cite{user_grouping_Shanmugam}, strong-edge colorings of bipartite graphs \cite{strongedgecoloringofbipartitegraphs}, a new combinatorial structure called placement-delivery array (PDA)\cite{YCTCPDA} and further such improved constructions \cite{CJYT,Cheng_Linear_TCOMM_HammingDistance}, hypergraph-based schemes \cite{SZG}, induced matchings of graphs \cite{STD}, resolvable designs \cite{TaR}, combinatorial designs \cite{Jian_PDA_CombiDesigns_BIBD_havetolook}, orthogonal arrays \cite{Cheng_OrthoArrays_SmallSubp_FlexibleMemory,Cheng_TIT_Framework_havetolookwithin}, projective geometry based constructions \cite{Kr_LineGraphs,Suthan_etal_Subexp_ProjGeo}, and even constructions that avoid file-splitting \cite{SLB}. This list is arguably not exhaustive and continues to grow. While most of this literature discusses the paradigm of coded caching, many of these can be lifted to the MapReduce setting via simple techniques \cite{PDAApp,QYDC,Q.Yan_PDA_CDC_stragglers}. Recently, techniques for constructing  new PDAs from existing ones by cleverly `lifting' existing  PDAs have also been proposed  \cite{Aravind_LiftingConstruction,Wang_Cheng_CartesianProduct}. In any case, there is indeed a tradeoff involving the three central performance parameters in any coded caching scheme namely the cache-fraction $M/N$, the rate $R$, and the subpacketization level $F$, as shown by the lower bounds on the rate of coded caching schemes in \cite{WTP,cheng2017coded,Suthan_etal_Subexp_ProjGeo}. The lower bound in \cite{cheng2017coded} holds for PDA-based schemes, whereas the bounds in \cite{WTP,Suthan_etal_Subexp_ProjGeo} are information-theoretic in nature and hold for all (even non-linear) coded caching schemes. Specifically, these lower bounds suggest that only two of these three parameters can be simultaneously reduced, for a reasonable number of clients. 

Many of the coded caching schemes in literature mentioned above focus on constructing low rate, low cache-fraction schemes, that result in high subpacketization level (often exponential in $K$, the number of users). This seems a natural choice, as in practice, we expect that the local caches are much smaller in size than the file library itself, as the file library can be extremely large (for instance, in a content delivery network, to offer the users a large number of choices in the content to download). However, there can be network scenarios where the network bandwidth may be the primary parameter of concern while the local storage may be able to cache a greater fraction of the complete file library. Some examples of this could be a network that serves popular videos to a small audience that have access to sufficient local storage, or a network in which the bandwidth is prohibitively costly or rare, while the cache itself may be quite inexpensive, such as communication in critical environments. Some existing coded schemes in the literature are defined for flexible memory sizes and hence exist for large cache fractions also. In this work, we present a number of new coded caching schemes, for the high cache-fraction, low subpacketization, low rate regime. We present both asymptotic (as $K$ increases) and numerical comparisons with some important existing schemes and existing lower bounds, showing the advantages of our schemes in a number of cases.


We formulate our constructions using a binary matrix model for coded caching. In Section \ref{sec:binarymatrices_cc_model}, we introduce the concept of using a constant row-weight binary matrix for describing the coded caching scheme. We call these as \textit{caching matrices}. The `1's in the binary matrix indicate uncached subfiles in the users. Identity submatrices of the caching matrix correspond to transmissions which enable the clients (involved in any transmission) to decode precisely one missing subfile each from that transmission. Thus, `covering'  the `1's in the caching matrix using identity submatrices provides a valid delivery scheme. The framework we present using binary matrices is essentially equivalent to the PDA framework developed in \cite{YCTCPDA}. However, the advantage is that this viewpoint opens up a much larger space, viz. the space of all constant row-weight binary matrices, for searching for good caching schemes.  

Following this, we use the binary matrix model for constructing novel caching schemes derived from a variety of combinatorial designs and their $q$-analogs, also called as \textit{subspace designs}. Towards that end, Section \ref{sec:designsbackground} describes important terminologies related to combinatorial designs and their $q$-analogs. In Sections \ref{sec:BIBD}-\ref{sec:subspace_designs_scheme}, we provide the details of the construction of caching matrices using different combinatorial and subspace designs. In particular, we employ Steiner systems ($t$-designs with special properties), balanced incomplete block designs, transversal designs, and $q$-analogs of Steiner systems, to construct caching matrices. When we employ existing designs from combinatorics literature to these constructions, the caching schemes which we get demand a low uncached fraction, i.e., $1-\frac{M}{N}=\Theta(\frac{1}{K^i}), i=\frac{1}{2},1$. This is a disadvantage. However, this disadvantage is traded off by a deep reduction in the rate as well as the subpacketization levels, with the schemes achieving a constant rate or even lower, with subpacketization levels being only at most linear in $K$. Section \ref{sec:summary_codedcaching} summarizes all our constructions and discusses their asymptotics wherever applicable, in terms of increasing number of clients.

We further show that we can utilize the designed binary matrices as schemes for the MapReduce framework as well. After a brief review of the coded MapReduce setup in Section \ref{CDR:systemmodel}, we introduce the notion of computing matrices for distributed computing in Section \ref{binarymatricesforcomputing}. For this setup, we propose a new simple delivery scheme in Section \ref{lowcomplexitydatashufflingsubsection}. We interpret the optimal coded MapReduce scheme shown in \cite{RDC} as a binary matrix based scheme, and show that the load achieved by our data-shuffling scheme for the same is strictly less than twice that of the optimal load. However, our scheme has lesser complexity in the data shuffling phase, as it avoids the splitting the IVAs further into smaller packets. We discuss these in Section \ref{advantagessubsubsection}. We show that  binary matrix constructions presented in Sections \ref{sec:BIBD}-\ref{sec:subspace_designs_scheme} result in coded MapReduce schemes, and show their parameters. Compared to the optimal scheme, these schemes give a larger communication load,  but have very low file complexity (Section \ref{combinatorialdesigns}). By default, our new scheme does not ensure \textit{communication load balancing}, i.e., not all servers participate in the transmissions during the data-shuffling phase; but such load balancing can be achieved by finding perfect matchings on an appropriately defined graph (Section \ref{loadbalancingsubsec}). Interestingly, the load imbalanced feature of our raw scheme can be exploited in order to protect against stragglers. In Section \ref{fullstragglersec}, we extend our scheme to distributed computing with stragglers. In the full straggler scenario, we assume that some nodes (numbering up to some threshold based on the properties of the computing matrix) are unavailable or failed. For this full straggler model, which we consider in Section \ref{DC_FullStragglers}, we show a revised scheme in which the communication load is increased by a factor that depends only on the increase in the number of functions reduced per node. For a fixed number of partial stragglers (nodes not fully unavailable but only slow), we show in Section \ref{DC_partialStragglers} that our scheme requires only slight modification to work, without any additional communication load.

In Section \ref{sec:numerical}, we present numerical results which compare our designs-based schemes for coded caching and for distributed computing, with some baseline schemes in each of these frameworks. For coded caching, we match the number of clients cache fractions (between $0.75$ to $0.9$) between our schemes and these baseline schemes as closely as possible, and compare the subpacketization and the rate. These are shown in Tables \ref{comparison_man} and \ref{comparison_sec}. In the case of MapReduce, we present comparisons between the new schemes presented in this work and some baseline schemes for the non-straggler and the straggler scenarios. These are shown in Tables \ref{comparison_straggler} and \ref{comparison_qys}. In general, our schemes are shown to have advantages when it comes to the subpacketization level, trading this advantage for a marginal increase in the communication load. We summarize our contributions in Section \ref{discussion} with some promising directions for future work.


\textit{Notations and Terminology: }
For any positive integer $N$, we denote by $[N]$ the set $\{1,\hdots,N\}$. For a set ${\cal X}$ and some positive integer $t\leq {|\cal X|}$, we denote the set of all $t$-sized subsets of ${\cal X}$ by $\binom{{\cal X}}{t}$. The binomial coefficient is denoted by $\binom{n}{r} \triangleq \frac{n!}{r!(n-r)!}$ for $n\geq r \geq 0$. For a matrix $A$ whose rows are indexed by a finite set ${\cal R}$ and columns are indexed by a finite set ${\cal C}$, the element in the $r^{th}$ row  $(r \in {\cal R})$ and $l^{th}$ column  $(l \in {\cal C})$ is denoted as $A(r,l)$. For sets $A,B,$ $A\backslash B$ denotes the elements in $A$ but not in $B$. For some element $i$, we also denote $A\backslash \{i\}$ by $A\backslash i$. For $j\in\{0,1,\hdots,(k-1)\}$, we denote $(j+1)mod~k$ by $(j {\oplus_k}1)$. For nonnegative integers $k\leq v$ and a prime power $q$, the notation $\gbinom{v}{k}$ denotes the Gaussian binomial coefficient given by 
$\gbinom{v}{k} \triangleq \frac{(q^v-1)\hdots(q^{v-k+1}-1)}{(q^k-1)\hdots(q-1)}.$ Note that we suppress $q$ in this notation for convenience. This number $\gbinom{v}{k}$ also is equal to the number of subspaces of dimension $k$ in any $v$ dimensional vector space over $\fq$, the finite field with $q$ elements. 



\section{Binary Matrices and Coded Caching}
\label{sec:binarymatrices_cc_model}
In this section, we describe the idea of an identity submatrix cover of a binary matrix, and show that a binary matrix with constant row weight and an identity submatrix cover results in a coded caching scheme. 

For the sake of formality, we identify some simple quantities and assign them  some terminology. We refer to a matrix with entries from $\{0,1\}$ as a binary matrix. Formally, a submatrix of a matrix $A$ can be specified by a subset of the row indices and a subset of column indices which we use to index the rows and columns of $B$ respectively. We refer to a $l\times l$ submatrix $B$ of matrix $A$ as an \textit{identity submatrix} of matrix $A$ of size $l$, if the columns of $B$ correspond to the identity matrix of size $l$ permuted in some way. For a binary matrix $A$, a non-zero entry $A(i,j)=1$ in the row $i$ and column $j$ is said to be \textit {covered} by the identity submatrix $B$ if $i$ and $j$ correspond to some row and column index of $B$ respectively. Two distinct identity submatrices $B_1$ and $B_2$ are said to \textit{overlap} when some non-zero entry in matrix $A$ is covered by both $B_1$ and $B_2$. 
\begin{definition}[Non-Overlapping Identity Submatrix Cover]
Consider a set $\mathfrak {C}=\{C_1,...,C_S\}$ consisting of $S$ identity submatrices of matrix $A$ such that each non-zero element in $A$ is covered by atleast one $C_i\in \mathfrak {C}$. Then, $\mathfrak {C}$ is called an identity submatrix cover of $C$.  Further, if no two identity submatrices in $\mathfrak{C}$ are overlapping, we call $\mathfrak{C}$ a non-overlapping identity submatrix cover. 
\end{definition}

We now recall the system model for coded caching as in \cite{MaN}. In the coded caching system in \cite{MaN}, we have a set of $K$ clients indexed by some set ${\cal U}$ of size $K$ and a single server.  There is a library of files at the server, consisting of $N$ files of the same size, which are denoted as $W_i : \forall i \in [N]$. Each client possesses local storage (called cache) which can store up to a fraction $M/N$ (called the \textit{cache fraction}) of the library, for some $M\in\{1,\hdots,N\}$.  The clients are connected to the server through an error free shared link. Each file is partitioned into $F$ non-overlapping subfiles of the same size, where $F$ is known as the \textit{subpacketization level}. The subfiles of $W_i$ are labelled as $W_{i,f}: \forall f\in {\cal F}$, ($W_{i,f}$ is assumed to take values from an abelian group) where ${\cal F}$ is a set of size $F$. The centralized coded caching framework consists of two phases: the \textit{placement phase} and the \textit{delivery phase}. The placement phase occurs during non-peak hours. In the placement phase, the communication channel is utilized so that the caches at the clients are fully populated by storing some subfiles of the files in each of them. The delivery phase occurs during the peak-hours. In the delivery phase (during peak hours), user $u$ demands a specific file $W_{d_u}$, where $d_u\in[N]$. The delivery scheme consists of sending coded transmissions of the subfiles of $W_{d_u}:u\in{\cal U}$ so that the demands are satisfied. The rate $R$ of the coded caching scheme is defined as the ratio of the number of bits transmitted by the server to the size of each file (in bits). When the delivery scheme is linear, i.e., each server transmission is a linear combination of the subfiles of $W_{d_u}:u\in{\cal U}$, then the rate can be calculated as 
\[
\text{Rate}~R =  \small \frac{\text{Number of bits transmitted by server}}{\text{Size of the file in bits}}=\small \frac{\text{Number of transmissions in the delivery phase}}{F}.
\]
We will present new coded caching schemes for this setup initiated by \cite{MaN}. For the purpose of presenting our coded caching schemes, we define \textit{caching matrices}, which essentially capture the caching phase of the coded caching scheme. 
\begin{definition}
[Caching Matrix] Consider a binary matrix $C$ with rows indexed by a $K$-sized set ${\cal U}$ and columns indexed by a $F$-sized set ${\cal F}$  such that the number of $1$'s in each row is constant (say $Z$).
Then the matrix $C$ defines a caching scheme with $K$ users (indexed by ${\cal U}$), subpacketization $F$ (indexed by ${\cal F}$) and uncached fraction $(1-\frac{M}{N}) = \frac{Z}{F}$ as follows:
\begin{itemize}
\item User $u\in {\cal U}$ caches $W_{i,f}: \forall i \in [N]$ if $C(u,f) = 0$ and does not cache it if $C(u,f) = 1$.
\end{itemize}
We then call the matrix $C$ as a $(K, F, (1-\frac{M}{N}))$-caching matrix.
\end{definition}

A subfile $W_{i,f}$ is said to be \textit{missing} at a user $u$ if it is not available at its cache. In order to construct a transmission scheme, we first describe one transmission based on the above described matrix based caching scheme, which will serve a number of users. 

\begin{lemma}
\label{transmissions_def}
Consider an identity submatrix of $C$ given by rows $\{u_1,u_2,..,u_l: u_i \in {\cal U}\}$ and columns $\{f_1,f_2,..,f_l: f_i \in {\cal F}\}$, such that $C(u_i,f_i)=1, \forall i \in [l]$, while $C(u_i,f_j)=0, \forall i,j \in [l]$ where $i\neq j$. For each $i \in [l]$, the subfile $W_{d_{u_i},f_i}$ is not available at user $u_i$ and can be decoded from the transmission $\sum_{i=1}^l W_{d_{u_i},f_i}$.
\end{lemma}

\begin{IEEEproof}
By definition of identity submatrix, for each $i \in [l]$ the subfile $W_{d_{u_i},f_i}$ is not available at user $u_i$ but is available at the users $\{u_1,u_2,..,u_l\} \backslash u_i$. Hence each user $u_i : i \in [l]$ can decode the subfile $W_{d_{u_i},f_i}$ which is not available at its cache from the transmission $\sum_{i=1}^l W_{d_{u_i},f_i}$.
\end{IEEEproof}
We now describe how an identity submatrix cover of $C$ is used to form a transmission scheme.
\begin{theorem}
\label{rate def}
Consider an identity submatrix cover $\mathfrak{C}= \{C_1,C_2,..,C_S\}$ of a caching matrix $C$. Then the transmission corresponding to $C_i:i \in [S]$ according to Lemma \ref{transmissions_def}, is a valid transmission scheme (i.e the scheme  satisfies all the user demands) for the caching scheme defined by $C$ and the rate of the transmission scheme, $R = \frac{S}{F}$.
\end{theorem}
\begin{IEEEproof}
Pick some arbitrary missing subfile $ W_{d_{u},f}$ of user $u$. Then $C(u,f) =1$ and this entry of $C$ will be covered by at least one of the identity submatrices, say $C_i$ in $\mathfrak{C}$ since $\mathfrak{C}$ is an identity submatrix cover of $C$. The transmission corresponding to the identity submatrix $C_i$ given by Lemma \ref{transmissions_def} will ensure that the subfile $ W_{d_u,f}$ will be decoded by the corresponding user $u$ where it is missing. Hence, the transmission corresponding to $C_i \in \mathfrak{C}$ enables decoding of any arbitrary missing subfile. Since the number of identity submatrices  in $\mathfrak{C}$ is $S$, the rate of the transmission scheme is, $R=\frac{S}{F}$.
\end{IEEEproof}

\section{Background on Combinatorial designs}
\label{sec:designsbackground}
In the previous section, we have developed a binary matrix model for the caching problem. In the upcoming sections, we will use combinatorial and subspace designs to construct caching matrices. For that purpose, we first review some of the basic definitions related to combinatorial designs and their constructions. For more details regarding the combinatorial designs, the reader is referred to \cite{Drs,col}. Later, we review some relevant aspects of subspace designs, which are the $q$-analogs of combinatorial designs. 
\begin{definition}[Design $({\cal X},{\cal A})$] A design is a pair $({\cal X},{\cal A})$ such that the following properties are
satisfied:
\newline (D1). ${\cal X}$ is a set of elements called points, and
\newline   (D2). ${\cal A}$ is a collection (i.e., multiset) of nonempty subsets of ${\cal X}$ called blocks.
\end{definition}
We now define $t$-designs.

\begin{definition}[$t$-designs]
Let $v, k, \lambda ,$ and $t$ be positive integers such that $v>k \geq t$. A $t$-$(v, k, \lambda)$-design (or simply $t$-design) is a design $(\cal X, A)$ such that the following properties are satisfied: \newline (T1). ${|\cal X|} = v$,
 \newline   (T2). Each block contains exactly $k$ points, and
  \newline   (T3). Every set of $t$ distinct points is contained in exactly $\lambda$ blocks.
  \end{definition}
Consider a nonempty $Y\subseteq {\cal X}$ such that $|Y| = s \leq t$. Then there are exactly
\begin{align}
\label{s_points}
{\lambda_s}  = \lambda \frac{\binom{v-s}{t-s}}{\binom{k-s}{t-s}}
\end{align}
blocks in ${\cal A}$ that contain all the points in $Y$. It can also be shown that $b={\lambda_0}=\lambda \dfrac{\binom{v}{t}}{\binom{k}{t}}$ is the number of blocks in $t$-designs. 

\begin{example}
\label{t_ex}
[Parametrized Constructions]
A $t$-design with $\lambda =1$ (i.e  $t$-$(v,k,1)$ design) is called a Steiner system and its existence is discussed in \cite{keevash}. A construction of Steiner system for $t=3$ and $t=4$ is presented in \cite{pappu}. Other general constructions for Steiner systems can be found in \cite{col}. Here we use a specific construction.

\begin{itemize}
    \item A construction of Steiner system with parameters $t=3,~ v=q^{2}+1,~ k=q+1$ is presented in \cite{Drs}, where $q$ is a prime power such that $q \geq 2$.
\end{itemize}
\end{example}

In the following examples and some others in this paper, we drop the parentheses and the commas in writing the blocks explicitly (for instance, block $\left\{l,m,n \right\}$ is written as $lmn$).
\begin{example}
\label{ex4}
Consider the following set $\cal X$ and an associated collection $\cal A$ of its subsets. 
${\cal X} = \left\{1,2,3,4,5,6,7,8 \right\}$ and ${\cal A} = \{ 1256,3478,1357,2468,1458,2367,
1234,5678,1278,3456,1368,2457,1467,2358 \}$.  It is not difficult to check that each $3$-sized subset of $\cal X$ is present in exactly one block (a subset in $\cal A$). Thus, this is a $3$-$(8,4,1)$ design (Steiner system). 
\end{example}

\begin{definition}[Balanced Incomplete Block Design] 
$t$-Designs with $t=2$ are called Balanced Incomplete Block Designs, (BIBD) denoted as ($v$, $k$, $\lambda$)-BIBD.
\end{definition}

By (\ref{s_points}) it follows that every ($v$, $k$, $\lambda$)-BIBD has exactly $b = {\frac {vr}{k}}$ blocks and, the number of blocks containing each point is exactly
\begin{align}
\label{r_points}
r = {\frac {\lambda(v-1)}{k-1}}.
\end{align}

\begin{example}
\label{ex1}
Let ${\cal X} = \left\{1,2,3,4,5,6,7,8,9 \right\}$ and ${\cal A} = \{ 357, 123, 456, 789, 147, 258, 369, 159, 267, 348, 168, 249 \}.$ The number of blocks is $b=12$ and each element in ${\cal X}$ occurs exactly in $r=4$ blocks. Also note that every pair of elements occurs in exactly one block in $\cal A$. Thus, $(\cal X,\cal A)$ is an example of a (9,3,1)-BIBD.
\end{example}
We now define Symmetric BIBDs.
\begin{definition}[Symmetric BIBD]
A $(v,k,\lambda)$-BIBD in which $b = v$ (or, equivalently, $r = k$) is called a symmetric BIBD.
\end{definition}
As stated in \cite{Drs}, for any two blocks $A_1, A_2 \in {\cal A}$ in a symmetric BIBD 
 \begin{align}
\label{lambda}
|A_1 \cap A_2| = \lambda.
\end{align}

\begin{example}
\label{bibd_const}
[Parametrized Constructions]
Some constructions of BIBD known in literature are given below:
\begin{itemize}
    \item Symmetric BIBDs with parameters $v= n^{2}+n+1$, $k= n+1$, $\lambda = 1$ are constructed in \cite{Drs} using a projective plane of order $n$, where $n$ is a prime power such that $n \geq 2$.
    
    \item BIBDs with parameters $v= n^{2}$, $k= n$, $\lambda = 1$ are constructed in  \cite{Drs} using an affine plane of order $n$ where, $n$ is a prime power such that $n \geq 2$.
    
    \item A construction of symmetric BIBDs using affine resolvable BIBDs is presented in \cite{arbibd}.
\end{itemize}
\end{example}

We now define Transversal Designs.

\begin{definition}[Transversal Designs] A transversal design of order or groupsize $n$, blocksize $k$, and index $\lambda$, denoted as $TD_{\lambda}(k, n)$, is a triple $( \cal X, \cal G, \cal B)$, where \newline (TD1). $ {\cal X}$ is a set of $kn$ elements.
\newline     (TD2).  $\cal G$ is a partition of ${\cal X}$ into $k$ sets (the groups), each of size $n$.
\newline     (TD3).  $\cal B$ is a collection of $k$-sized subsets of $ {\cal X}$ (the blocks).
\newline     (TD4).  Every pair of elements from ${\cal X}$ is contained either in exactly one group or in exactly $\lambda$ blocks, but not both.
\end{definition}
Transversal designs in which $\lambda = 1$, are denoted by TD$(k, n)$. From the above properties, we see that  $|\cal B|$= $n^{2}$ and each element of $ {\cal X}$ occurs in $n$ blocks for $\lambda=1$ \cite{han}.
\begin{example}
\label{ex5}
A TD$(4,3)$ design is given as follows. Consider sets ${\cal X} =\{1,2,3,4,5,6,7,8,9,10,11,12   \} $, ${\cal G} = \{ \{1,2,3 \},\{4,5,6 \}, \{7,8,9 \},\{10,11,12 \} \}$, and \newline
$\scriptsize {\cal B} = \{ \left\{1,4,7,10 \right\}, \left\{1,5,8,11 \right\}, \left\{1,6,9,12 \right\}, \left\{2,4,9,11 \right\}, 
\left\{2,5,7,12 \right\}, \left\{2,6,8,10 \right\}, \left\{3,4,8,12 \right\}, \left\{3,5,9,10 \right\}, \newline \left\{3,6,7,11 \right\} \}.$ 
It can be checked that every pair of elements in $\cal X$ is present in exactly one group or exactly one block, but not both. 
\end{example}
\begin{example}
\label{ex_td1}
[Parametrized Constructions] Some constructions of TD$(k,n)$ known in literature \cite{Drs} are as follows:
\begin{itemize}
    \item A transversal design with parameters $\lambda =1, k=q, n=q$ can be constructed using orthogonal arrays, where $q$ is a prime power such that $q\geq 2$.
     \item A transversal design with parameters $\lambda =1, k=q+1, n=q$ can be constructed using orthogonal arrays, where $q$ is a prime power such that $q\geq 2$.
     
\end{itemize}
\end{example}

The final combinatorial structure we consider in our work is the notion of subspace designs, which are $q$-analogs of combinatorial designs. We recall some basic definitions regarding these objects here. More details regarding these objects can be found in \cite{Braun2018,braun_etzion_östergård_vardy_wassermann_2016,etzion2009q}. 
\begin{definition}[Subspace Designs] Let ${\cal V}$ be a vector space over the finite field $\fq$ of dimension $v$. Let the subspaces with dimension $k$ be called as $k$-dim subspaces. The q-analog of a design is defined as follows. Let $0 \leq t \leq k \leq v$ be integers and $\lambda$ be a non-negative integer. A pair  ${\cal D} = ({\cal V},{\cal A})$, where ${\cal A}$ is s a collection of $k$-dim subspaces (blocks) of ${\cal V}$, is called a $t$-$(v, k, \lambda)_q$-subspace design on ${\cal V}$ if each $t$-dim subspace of ${\cal V}$ is contained in exactly $\lambda$ blocks.
\end{definition}
A $t$-$(v,k,1)$ subspace design is also referred to as a $q$-analog of an equivalent Steiner system. We now recount some known constructions of subspace designs. 
\begin{example}
\label{subspacedesigns_examples}
Few constructions of subspace designs known from literature are recollected here. 
\begin{enumerate}
    \item For any $0\leq t\leq k\leq v$, and any $q$ being a prime power, the collection of all $k$-dimensional subspaces of $\fq^v$ forms a $t-(v,k,\gbinom{v-t}{k-t})_q$ subspace design. Specifically, when $t=k$, we get a $k-(v,k,1)_q$ subspace design.  
    \item For any prime power $q$, it was shown in \cite{Braun2018} that there exists a $1-(v, k, 1)_q$ design if and only if $k$ divides $v$. 
    \item In \cite{Suzuki1990,Suzuki1992}, a construction of nontrivial $2-(v,3,q^2+q+1)_q$ designs was presented, for all $q$ being a prime power and for all $v\geq 7$ such that $v$ and the integer $24$ are coprime. 
    \item The authors in \cite{Fazeli_Lovett_Vardy_Nontrivialtdesigns} showed the existence of subspace designs over $\fq$ for any $t$ and any $k> 12(t+1)$, when $n$ is sufficiently large. 
    \item Many other individual constructions for specific parameters are available in \cite{Braun2018}, some of which we use in this work to present numerical examples of the coded caching schemes obtained from subspace designs.
\end{enumerate}
\end{example}

We will use in some constructions the following idea of the \textit{incidence matrix} of a design.
\begin{definition}[Incidence Matrix] Let ($\cal X, \cal A $) be a design where ${\cal X}$ = $\left\{x_1,...,x_v \right\}$ and $\cal A$ =
$\left\{A_1,...,A_b \right\}$. The incidence matrix of ($\cal X, \cal A $) is the $v\times b$ binary matrix $M =(M(i,j))$ defined by the rule

$M(i,j) =
    \begin{cases}
      1, & \text{if}\ x_i\in A_j , \\
      0, & \text{if}\ x_i\not\in A_j. \\
    \end{cases}$
\end{definition}
\begin{example}
\label{ex_5}
The incidence matrix for $(7,3,1)$-BIBD given by ${\cal X} = \left\{1,2,3,4,5,6,7 \right\}$
 and ${\cal A} = \{127,145,136,467, \newline 256,357,234 \}$ is given below.  
\renewcommand{\kbldelim}{(}
\renewcommand{\kbrdelim}{)}
\[\small
  \text{$C$} = \kbordermatrix{
    & 127 & 145 & 136 & 467 & 256 & 357 & 234 \\
    1 & 1 & 1 & 1 & 0 & 0 & 0 & 0\\
    2 & 1 & 0 & 0 & 0 & 1 & 0 & 1\\
    3 & 0 & 0 & 1 & 0 & 0 & 1 & 1\\
    4 & 0 & 1 & 0 & 1 & 0 & 0 & 1\\
    5 & 0 & 1 & 0 & 0 & 1 & 1 & 0\\
    6 & 0 & 0 & 1 & 1 & 1 & 0 & 0\\
    7 & 1 & 0 & 0 & 1 & 0 & 1 & 0
  }.
\]

\end{example}

\section{Summary of New Coded Caching Schemes from Designs}
\label{sec:summary_codedcaching}
Table \ref{tab1} summarizes all the caching parameters related to the coded caching schemes to be constructed from the various designs in the forthcoming sections. The parameters $v,k,t$ are based on those of the designs using which they are constructed. Table \ref{tab1} also lists in the last two columns the (optimal) rate $R^*$ of the Ali-Niesen scheme \cite{MaN} and the corresponding subpacketization level $F^*$ (which was shown to be optimal across all PDA-based schemes which achieve the optimal rate $R^*$ in \cite{cheng2017coded}).

Applying the results of Table \ref{tab1} to the parameterized constructions of designs as given in Section \ref{sec:designsbackground}, we get the following results regarding the specific constructions obtained in this paper listed in Section \ref{subsec:specificconstructions}. Numerical results comparing many of these specific constructions to existing baseline schemes are shown in Section \ref{sec:numerical}. 

 





\begin{table*}[ht]
\centering
\begin{tabular}{|M{2.3 cm}|c|c|c|c|c|c|}
\hline
Combinatorial Design $ (\lambda = 1)$ &$(1-\frac{M}{N})$ & $K$ & $F$ & $R$ & $R^{*}( = (\frac{K(1-\frac{M}{N})}{1+ K\frac{M}{N}}))$ & $F^{*} (= \binom{K}{K\frac{M}{N}})$\\
\hline
BIBD   & $\frac{k}{v} $& $v$ & $\frac{v(v-1)}{k(k-1)}$ & $\frac{k(k-1)}{v}$ & $\frac{k}{1+v-k}$ & $\binom{v}{k}$ \\ 
\hline


t-design  & $\frac{\binom{k}{t}(v-t+1)}{\binom{v}{t}k}$& $\binom{v}{t-1}$ & $\frac{\binom{v}{t}k}{\binom{k}{t}}$ & $\frac{\binom{k-1}{t-1} \binom{k}{t}v}{\binom{v}{t}k}$& $\frac{\binom{k-1}{t-1}}{1 + \binom{v}{t-1} - \binom{k-1}{t-1}}$ & $\binom{\binom{v}{t-1}}{\binom{k-1}{t-1}}$ \\ 
\hline


Transversal Design  &$\frac{1}{n} $& $n^2$ & $kn$ & 1 & $\frac{n}{1-n + n^2}$ & $\binom{n^2}{n}$ \\ 
\hline

Subspace Design & $\frac{\gbinom{v-t+1}{1}q^{t-1}}{F}$ & $\gbinom{v}{t-1}$ & $\frac{\gbinom{v}{t}\gbinom{k}{1}}{\gbinom{k}{t}}$ & $\frac{\gbinom{k-1}{t-1} \gbinom{v}{1}q^{t-1}}{F}$ & (too big to fit) & (too big to fit) \\
\hline

\end{tabular}
\caption{Parameters of Coded Design-based coded caching schemes of this paper with lower bounds of \cite{MaN}. Certain entries which do not fit in the table can be easily computed using the expression of $R^*$ and $F^*$ on the top row.}
\label{tab1}. 
\hrule
\end{table*}
\subsection{Specific Constructions of Coded Caching Schemes from Existing Designs}
\label{subsec:specificconstructions}
\subsubsection{BIBDs}
\label{sc_bibd}
The parameters of the transmission scheme (described in Section \ref{sec:BIBD}) for the constructions described in Example \ref{bibd_const} are as follows:
\begin{enumerate}[label=\alph*)]
  \item Symmetric BIBDs with parameters $v= n^{2}+n+1$, $k= n+1$, $\lambda = 1$ will give a coded caching scheme with parameters $F=n^{2}+n+1$, $K=n^{2}+n+1$, Rate $= 1$, $(1-\frac{M}{N}) = \frac{n+1}{n^{2}+n+1}$. 
    
    \item BIBDs with parameters $v= n^{2}$, $k= n$, $\lambda = 1$ will give a coded caching scheme with parameters $F=n^{2}+n$, $K=n^{2}$, Rate $= \frac{n}{n+1}$, $(1-\frac{M}{N}) = \frac{1}{n}$.
\end{enumerate}

Note that for the above two schemes, we have $F=O(K)$, and $1-\frac{M}{N}=\Theta(\frac{1}{\sqrt{K}}),$ while $R\leq 1.$
\subsubsection{Steiner systems}
\label{sc_steiner}
For the constructions described in Example \ref{t_ex}, the parameters of the transmission scheme presented in Section \ref{sec:t-section} are as follows.
\begin{enumerate}[label=\alph*)]
\item $t$-designs with parameters $t=3,~v=q^{2}+1,~k=q+1, ~\lambda=1$,  will give a coded caching scheme with parameters $F=(q^{2}+1)(q+1)$, $K=\frac{(q^{2}+1)q^{2}}{2}$, Rate $= \frac{(q-1)}{2(q+1)}$, $(1-\frac{M}{N}) = \frac{(q-1)}{q(q^{2}+1)}$.

\end{enumerate}

We note that for the above construction of the coded caching scheme, we have $F=O(K^{\frac{3}{4}}),$ $1-\frac{M}{N}=\Theta(\frac{1}{\sqrt{K}})$ and $R\leq 1.$



\subsubsection{Transversal Designs}
\label{sc_tranversal}
The parameters of the transmission scheme (presented in Section \ref{sec:transversal}) for the constructions described in Example \ref{ex_td1} are as follows:

\begin{enumerate}[label=\alph*)]

     \item A transversal design with parameters $\lambda =1,~ k=q,~ n=q$, will give a coded caching scheme with parameters $F=q^{2}$, $K=q^{2}$, Rate $= 1$, $(1-\frac{M}{N}) = \frac{1}{q}$.
     
     \item A transversal design with parameters $\lambda =1,~ k=q+1,~ n=q$, will give a coded caching scheme with parameters $F=q^{2}+q$, $K=q^{2}$, Rate $= 1$, $(1-\frac{M}{N}) = \frac{1}{q}$.
\end{enumerate}


For both of the above constructions, we have $F=O(K)$ and $(1-\frac{M}{N})=O(\frac{1}{\sqrt{K}}).$

\subsubsection{Subspace Designs}
\label{sc_subspace} 
 For any prime power $q$, we have noted in Example \ref{subspacedesigns_examples} the existence of a $k-(v,k,1)_q$ subspace design (the blocks being the set of all $k$-dimensional spaces of the $v$-dimensional space over $\fq$). This leads to a coded caching scheme (shown in Section \ref{sec:subspace_designs_scheme}) with $K=\gbinom{v}{k-1}$, $F=\gbinom{v}{k}\gbinom{k}{1}$, $1-M/N=\frac{\gbinom{v-k+1}{1}q^{k-1}}{\gbinom{v}{k}\gbinom{k}{1}}$ and $R=\frac{\gbinom{v}{1}q^{k-1}}{\gbinom{v}{k}\gbinom{k}{1}}$.  Due to the interplay of these parameters, it is difficult to ascertain the asymptotics of this scheme. Interestingly, we see in our numerical results (Section \ref{sec:numerical}) that the caching schemes obtained from these subspace designs (and a few  examples of other subspace designs in literature) have some of the smallest cache-fractions (i.e., largest $1-M/N$ values) amongst the schemes designed in this work.

In the forthcoming sections, we provide constructions for caching schemes based on the above combinatorial and subspace designs. Each construction is contingent on the existence of the design of the considered type. In each such case, we define the caching matrix using the given design and obtain its parameters $K, F,$ and $(1-\frac{M}{N})$ based on the design parameters. We then define and prove an identity submatrix cover of the caching matrix based on the properties of the design. The format for the proof of the identity submatrix cover in each of the following constructions is the same, which we describe sequentially as follows:
\begin{enumerate}
    \item We describe a method to pick a submatrix of the caching matrix,  which we prove to be an identity submatrix in the following way.
    \begin{enumerate}
        \item We show that the submatrix has equal number of rows and columns. 
        \item We then show that each row and column of the submatrix has weight one. 
    \end{enumerate}
    \item We then show that the identity submatrices picked have no overlaps. 
    \item Finally, we show that all the $1$'s of the caching matrix are covered by the collection of identity submatrices, thus proving that the collection forms an identity submatrix cover. 
\end{enumerate}


\section{BIBD (with $\lambda =1$) based Coded Caching Scheme}
\label{sec:BIBD}

Consider a ($v,k,1$)-BIBD $({\cal X,A})$ with $|{\cal A}|=b$ being the number of blocks. We order the elements in ${\cal X}$ in some arbitrary way. For distinct $x,y \in {\cal X}$, we say that $x < y$ if $x$ comes before $y$ in the ordering of ${\cal X}$. Let the elements in block $A\in{\cal A}$ be denoted as $\{A(0),A(1),\hdots,A(k-1)\}$ where $A(0)<A(1)<..<A(k-1)$. Let $C$ denote the incidence matrix of ($v,k,1$)-BIBD.

\begin{remark}
\label{rem:bibdconstweight}
Note that each element in ${\cal X}$ occurs in $r=\frac{v-1}{k-1}$ blocks, by (\ref{r_points}) . Thus each row of the incidence matrix $C$ has weight $r$. Also, each block $A\in {\cal A}$ is of size $k$, thus column weight of the matrix $C$ is $k$. Hence the binary caching matrix constructed from BIBD has constant row and column weight.
\end{remark}

Now, $(1-\frac{M}{N}) = \frac{r}{b} = \frac{k}{v}$ (since $vr=bk$), and $F =b= \frac{v(v-1)}{k(k-1)}$.  Hence, the incidence matrix $C$ of design $({\cal X,A})$ will give a $\left(v,\frac{v(v-1)}{k(k-1)},\frac{k}{v}\right)$-caching matrix.
\par For $x \in {\cal X}$, let $B_x \triangleq\left\{A \in {\cal A} : x \in A\right\}$. Note that $|B_x| = r$, by the property of  BIBD with $\lambda=1$. In the next lemma, we describe a single identity submatrix of matrix $C$.


\begin{lemma}
\label{lemma:Cx_identity}
For any $x\in {\cal X}$, let us denote $B_x$ as $B_x=\left\{A_{1},...,A_{r}\right\}$ where $x=A_{i}(j_i) : j_i\in\{0,1,\hdots,k-1\}$ i.e $x$ is the $j_i^{th}$ element in the block $A_{i}$. Consider the submatrix $C_x$ of $C$ whose columns are indexed by $B_x$ and rows are indexed by $\left\{ A_{i}(j_{i} {\oplus_k} 1)  : i\in[r], A_{i} \in B_x \right\}$. Then $C_x$ is an identity submatrix of $C$ of size $r=\frac{v-1}{k-1}$.
\end{lemma}
\begin{IEEEproof}
Note that there are $r$ columns in submatrix $C_x$ of $C$. Now we show that there are $r$ rows. Due to the fact that $({\cal X,A})$ is a BIBD with $\lambda =1$, the elements $\{x,A_{i}(j_{i} {\oplus_k} 1)\}\subset A_{i}$ only and does not lie in any other $A_{i'}$ for any $i'\neq i$. Thus the elements $A_{i}(j_{i} {\oplus_k} 1)\neq A_{i'}(j_{i'} {\oplus_k} 1)$ for any $i'\neq i$. Hence, there are $r$ rows in $C_x$. Let us consider the row indexed by $A_{i}(j_{i} {\oplus_k} 1)$. Again by the property that $\lambda =1$, it holds in $C_x$ that the only column index corresponding to which there is a $1$ in the row indexed by $A_{i}(j_{i} {\oplus_k} 1)$ is $A_{i}$ and no other column. Thus each row of $C_x$ has a single entry $1$ in some column. Now consider an arbitrary column of $C_x$, say indexed by $A_{i}$. Since $A_{i'}(j_{i'} {\oplus_k} 1)\in A_{i}$ only if $i=i'$, thus the column $A_{i}$ has a $1$ only in the row $A_{i}(j_{i} {\oplus_k} 1)$. Hence, $C_x$ is an identity submatrix of $C$ of size $r$.
\end{IEEEproof}



In next two lemmas we will prove that there is no overlap between the identity submatrices $C_x : x \in {\cal X}$ and that these identity submatrices will cover all the entries where $C(x,A) =1$ in matrix $C$.
\begin{lemma}
\label{no_overlap}
For distinct $x_1,x_2\in{\cal X}$, there is no $x\in {\cal X}$, $A\in{\cal A}$ with $C(x,A)=1$ such that $C(x,A)$ is covered by both $C_{x_1}$ and $C_{x_2}$, where $C_{x_1}$ and $C_{x_2}$ are as defined in Lemma \ref{lemma:Cx_identity}.
\end{lemma}
\begin{IEEEproof}
Suppose $C(x,A)=1$ is covered by both $C_{x_1},C_{x_2}$. By the definition of $C_{x_1}$ and $C_{x_2}$, it implies that $x_1, x_2 \in A$. Let the element $x_1$ and $x_2$ be present in $j$ and $j'$ position of $A$ where $j \neq j'$. Therefore, $x= A(j {\oplus_k} 1) =x= A(j' {\oplus_k} 1)$, by our construction. But $j \neq j'$. This gives a contradiction. Hence, there is no $C(x,A)=1$ which is covered by both $C_{x_1}$ and $C_{x_2}$. 
\end{IEEEproof}

\begin{lemma}
\label{C_cover}
The set of matrices $\{C_x :x\in {\cal X}\}$ forms a non-overlapping identity submatrix cover of $C$.
\end{lemma}
\begin{IEEEproof}
The total number of $1$'s in matrix $C$ is equal to the product of the number of $1$'s in each row and the number of rows, and thus equal to $rv$. For each $x\in {\cal X}$ (note that ${|\cal X|}=v$), there exists an identity submatrix $C_x$. From Lemma \ref{lemma:Cx_identity} and \ref{no_overlap}, we see that each identity submatrix is of size $r$ and no two such identity submatrices have overlaps.  These identity submatrices will cover $vr$ number of $1$'s in matrix $C$ which is equal to total number of $1$'s in $C$. Hence, $\{C_x :x\in {\cal X}\}$ forms a non-overlapping identity submatrix cover of $C$.
\end{IEEEproof}
We thus have the following theorem summarizing the caching scheme.
\begin{theorem}
\label{main}
The incidence matrix of a ($v,k,1$)- BIBD forms a $\left(K=v,F=\frac{v(v-1)}{k(k-1)},(1-\frac{M}{N})=\frac{k}{v}\right)$ caching matrix. Further there is a transmission scheme with rate $R = \frac{k(k-1)}{v-1}$.
\end{theorem}
\begin{IEEEproof}
The parameters of the caching matrix $C$ have already been defined. By Lemma \ref{lemma:Cx_identity}, \ref{no_overlap} and \ref{C_cover}, we have an identity submatrix cover consisting of $v$ identity submatrices. Hence in Theorem \ref{rate def}, $S=v$ and rate $R = \frac{S}{F}= \frac{k(k-1)}{v-1}$.
\end{IEEEproof}


\begin{example}
\label{ex6}
Consider the $(7,3,1)$-BIBD as given in Example \ref{ex_5}. We describe the identity submatrix $C_1$ (as per Lemma \ref{lemma:Cx_identity}) corresponding to element `$1$' in ${\cal X}$. Element `$1$' is present in blocks $\{127, 145, 136\}$. Then the columns of identity submatrix matrix $C_1$ are indexed by $127, 145, 136$ and rows are indexed by $2, 4, 3$ (since $2,4,3$ are the next elements present after `$1$' in blocks $127, 145, 136$ respectively) of matrix $C$. Similarly, the identity submatrix $C_6$ corresponding to element `$6$' in ${\cal X}$ has column indices and row indices as $136,467,256$ and $1,7,2$ respectively in matrix $C$. In this manner, the submatrices $C_i: i \in 
{\cal X}$, gives us a rate $1$ transmission scheme.
\end{example}

\section{$t$-design based Coded Caching Scheme}
\label{sec:t-section}
We now describe a coded caching scheme via binary matrices arising out of 
$t$-designs. 

Let $({\cal X},{\cal A})$ denote a $t$-$(v,k,1)$ design. Let the blocks in this $t$-design be denoted as ${\cal A}=\{B_1,...,B_b\}$ where $b=\frac{\binom{v}{t}}{\binom{k}{t}}$. We construct a binary matrix $T$ as follows.
Let the rows of $T$ be indexed by all the $(t-1)$-sized subsets of $\cal X$. Let the columns be indexed by $\{(y,B):y\in B, B\in{\cal A}\}$. The number rows in matrix $T$ is $\binom{v}{t-1}$. The number of columns in matrix $T$ is $bk=\frac{{\binom{v}{t}k}}{\binom{k}{t}}$. For some $D\in\binom{{\cal X}}{t-1}$, 
the matrix $T= (T(D,(y,B)))$ is defined by the rule,
\newline $T(D,(y,B)) =
  \begin{cases}
   1, & \text{if}\ D\cup\{y\}\subset B, |D\cup\{y\}|=t \\
    0, & \text{otherwise}. \\
  \end{cases}$
  
\begin{remark}  
\label{rem:scheme1tdesignconstweight}
From the properties of the $t$-design, it is easy to calculate that the number of $1$'s in each row of $T$ is $\lambda_{t-1}\binom{k-t+1}{1}= (v-t+1)$. Also the number of $1$'s in each column is $\binom{k-1}{t-1}$. Hence, binary caching matrix $T$ is also a constant row and column weight matrix.
\end{remark}

Matrix $T$ gives a $\left(\binom{v}{t-1},\frac{{\binom{v}{t}k}}{\binom{k}{t}},\frac{{\binom{k}{t}(v-t+1)}}{{\binom{v}{t}}k}\right)$-caching matrix. In the next lemma we describe an identity submatrix of matrix $T$. Towards that end, we need to denote a few sets. For some $y \in {\cal X}$, define ${\cal B}_y \triangleq \{B \in {\cal A} : y \in B\}$, i.e the set of blocks containing $y$. By (\ref{s_points}), $|{\cal B}_y|=\lambda_1 = \frac{\binom{v-1}{t-1}}{\binom{k-1}{t-1}}$. Denote ${{\cal B}_y}$ by ${\cal B}_y=\{B_1,\hdots,B_{\lambda_1}\}$. For any $B_i$, denote by $\{D_{i,j} : j \in [\binom{k-1}{t-1}]\}$ the set $\binom{B_i\backslash y}{t-1}$, i.e the set of all $(t-1)$-sized subsets of $B_i\backslash y$.
\begin{lemma}
\label{lemma:t5_1}
For some $j\in[\binom{k-1}{t-1}] $ and $y \in {\cal X}$, consider the submatrix $T_{y,j}$ of $T$ whose rows are indexed by $\{D_{i,j} : \forall i \in [\lambda_1] \}$ ($D_{i,j}$ as defined above) and the columns are indexed by $\{(y, B_{i}): B_i \in {\cal B}_y\}$. Then $T_{y,j}$ is an identity submatrix of $T$ of size $\lambda_1=\frac{\binom{v-1}{t-1}}{\binom{k-1}{t-1}}$.   
\end{lemma}

\begin{IEEEproof}
Clearly, the number of columns in $T_{y,j}$ is $\lambda_1$. First note that the rows in $\{D_{i,j} : \forall i \in [\lambda_1]\}$ are all distinct, i.e $D_{i,j} \neq D_{{i'},j}$ for $i \neq {i'}$. If not, note that $D_{i,j} \cup y = D_{{i'},j} \cup y \in B_i \cap B_{i'}$. But this contradicts the fact that any $t$- sized subset of ${\cal X}$ occurs in only one block. Thus $|\{D_{i,j} : i \in [\lambda_1]\}|= \lambda_1=\frac{\binom{v-1}{t-1}}{\binom{k-1}{t-1}}$, invoking (\ref{s_points}), and thus $T_{y,j}$ is a square matrix of size $\lambda_1$. 

Now consider a row of $T_{y,j}$ indexed by $D_{i,j}$ for some particular $i \in [\lambda_1]$. Suppose a column indexed by $(y,B)$ for some $B \in {\cal B}_y$ has a $1$ in the row indexed by $D_{i,j}$. Then it means that $D_{i,j} \cup \{y\} \subset B$. But there is precisely one block $B$ such that $D_{i,j} \cup \{y\} \subset B$, which is precisely $B = B_i$ (as $\lambda=1$). Thus each row of $T_{y,j}$ has only one entry which is $1$. 

Now, consider a column of $T_{y,j}$ indexed by $(y,B_i)$ for some $B_i \in {\cal B}_y$. Suppose for some $D_{{i'},j}$, the row indexed by $D_{{i'},j}$ has $1$ in the column indexed by $(y,B_i)$. Then it must be that $D_{{i'},j} \subset B_i \backslash y$ and hence $D_{{i'},j} \cup \{y\} \subset B_i$. Once again, because of the property of $t$-design with $\lambda=1$, we have that $i = {i'}$ (else $D_{{i'},j} \cup \{y\} \in B_i \cap B_{i'}$, which is a contradiction). Hence, each column of $T_{y,j}$ has precisely only one entry that is $1$. This proves the lemma.   
\end{IEEEproof}
 
In the next two lemmas we will prove that there is no overlap between the identity submatrices and that these identity submatrices will cover all the entries where $T(D,(y,B)) =1$ in matrix $T$.
\begin{lemma}
\label{t5_2}
Any $T(D,(y,B)) =1$ such that $y\in {\cal X}, B\in{\cal A}, D\in \binom{\cal X}{t-1}$ will be covered by exactly one identity submatrix of $T$ (as defined in Lemma \ref{lemma:t5_1}).
\end{lemma}

\begin{IEEEproof}
Let $T(D,(y,B)) =1$ be covered by an identity submatrix $T_{{y'},j}$ of $T$.  As $T(D,(y,B)) =1$, we have that $D \cup \{ y\} \subset B$. By definition of $T_{{y'},j}$ in Lemma \ref{lemma:t5_1}, we must first have $y = {y'}$. Further it must be that $D \subset B_i \backslash y$ for some $B_i$ which contains $y$. Therefore, $\{y\} \cup D \subset B_i$, which means $B_i = B$ (as $\lambda=1$). Hence the unique transmission which covers $T(D,(y,B)) =1$ is $T_{y,j}$ where $j$ is such that $D=D_{i,j}$ is the unique $j^{th}$ set in $\binom{B \backslash y}{t-1}$.
\end{IEEEproof}

\begin{lemma}
\label{t5_3}
The set of matrices $\{T_{y,j}: \forall y\in {\cal X},~ \forall j\in [\binom{k-1}{t-1}]\}$ forms a non-overlapping identity submatrix cover of $T$.
\end{lemma}
\begin{IEEEproof}
The total number of $1$'s in $T$ is equal to the product of the number of $1$'s in each row and the number of rows, and thus equal to $(v-t+1)\binom{v}{t-1}=t \binom{v}{t}$. The size of each identity submatrix is $\frac{\binom{v-1}{t-1}}{\binom{k-1}{t-1}}$. Since we have an identity submatrix $T_{y,j}$ for each $j\in [\binom{k-1}{t-1}],~ y\in {\cal X}$, the number of identity submatrices $=\binom{k-1}{t-1}v$. Moreover, from Lemma \ref{t5_2}, there are no overlaps between the identity submatrices. Hence, the total number of $1$'s covered by all identity submatrices $= {\binom{k-1}{t-1}}{\frac{\binom{v-1}{t-1}v}{\binom{k-1}{t-1}}}=\binom{v-1}{t-1}v=t \binom{v}{t} $ which is equal to the total number of $1$'s in matrix $T$. Hence, $\{T_{y,j}: y\in {\cal X}, ~j\in [\binom{k-1}{t-1}]\}$ forms a non-overlapping identity submatrix cover of $T$.
\end{IEEEproof}
We thus have the below theorem summarizing the caching scheme.
\begin{theorem}
\label{T5_main}
The matrix $T$ of a $t$-($v,k,1$)-design forms a $\left(K=\binom{v}{t-1},F=\frac{{\binom{v}{t}k}}{\binom{k}{t}},(1-\frac{M}{N})=\frac{{\binom{k}{t}(v-t+1)}}{{\binom{v}{t}}k}\right)$-caching matrix. Further there is a transmission scheme with rate $R = \frac{{\binom{k-1}{t-1}}{\binom{k}{t}}v}{\binom{v}{t}k}$.
\end{theorem}

\begin{IEEEproof}
The parameters of the caching matrix $T$ have already been defined. By Lemma \ref{lemma:t5_1}, \ref{t5_2} and \ref{t5_3}, we have an identity submatrix cover consisting of $\binom{k-1}{t-1}v$ identity submatrices. Hence in Theorem \ref{rate def}, $S=\binom{k-1}{t-1}v$ and rate $R =\frac{S}{F}= \frac{{\binom{k-1}{t-1}}{\binom{k}{t}}v}{\binom{v}{t}k}$.
\end{IEEEproof}


\begin{example}
Consider the $3$-$(8,4,1)$ design as given in Example \ref{ex4}. We now describe some identity submatrices namely $T_{4,1},T_{4,2}$ and $T_{4,3}$. The set of blocks containing element `$4$' are denoted by ${\cal B}_4= \{3478,2468,1458,1234, \newline 3456, 2457,1467\}$. It is clear from our construction that the rows of $T$ are indexed by $(t-1)=2$-sized subsets of ${\cal X}$. The identity submatrices $T_{4,1},T_{4,2}$ and $T_{4,3}$ have rows indexed by $\{37,26,15,12,35,25,16\},~ \{38,28,18,13,36,27, \newline 17\}$ and $\{78,68,58,23,56,57,67\}$ respectively and columns indexed by $\{(4,B_i): B_i\in {\cal B}_4 \}$ of matrix $T$. In this manner we can obtain an identity submatrix cover of $T$ using the matrices $T_{y,j}: \forall y \in {\cal X},~ \forall j \in [\binom{k-1}{t-1}]$, which gives a transmission scheme with rate $=\frac{3}{7}$.
\end{example}

\section{Transversal Design based Coded Caching Scheme}
\label{sec:transversal}
Consider a TD($k,n$) transversal design $ ({\cal X}, {\cal G}, {\cal B})$ such that $k\geq n$. We order the group in some particular way. Thus, ${\cal G}=\{G_i : i \in \{0,1,\hdots,k-1\}\}$ where $|G_i|= n$, denotes the groups. Note that $|{\cal X}|=kn$. We see that by (TD$4$), for $\lambda =1$ each pair of elements from ${\cal X}$ is contained either in exactly one group in ${\cal G}$ or in one block in ${\cal B}$, but not both. Thus any two elements in a block must be from different groups. Therefore, for any block $A\in {\cal B}~(|A|=k)$, there is precisely one element  $A({G_i})$, such that $A({G_i}) \in B \cap G_i$. Thus, the elements in block $A\in {\cal B}$ can be ordered as $\{ A(G_0),\hdots,A(G_{k-1})\}$ where element $A(G_i), i \in \{0,1,\hdots,k-1\} $ is from group $G_i$. 

Let $C$ be the transpose of the incidence matrix of TD$(k,n)$. The number of rows and the number of columns in matrix $C$ are $n^2$ and $kn$ respectively.

\begin{remark}
\label{rem:constweighttransversaldesign}
It is easy to see that each row and column of matrix $C$ have  weights equal to $k$ and $n$ respectively. Hence, matrix $C$ is a constant row and column weight matrix.
\end{remark}

Matrix $C$ will give a ($n^{2},kn,\frac{1}{n}$)-caching matrix. Recall that each $x \in {\cal X}$ occurs in precisely $n$ blocks of ${\cal B}$, by the property of TD$(k,n)$. We now describe the lemma which gives an identity submatrix of matrix $C$. 


\begin{lemma}
\label{lemma:td_Cx_identity}
Define by ${\cal B}_x$ the set of blocks which contain $x$ and denote ${{\cal B}_x} = \{A_{1},\hdots,A_{n}\}$ where $x=A_{i}(G_{j_i}) : {j_i}\in \{0,1,\hdots,k-1\}$ i.e. $x$ is the ${j_i}$ element in block $A_i$. Consider the submatrix $C_x$ whose rows are indexed by ${\cal B}_x$ and columns are indexed by $\left\{A_{i}(G_{j_{i}{\oplus_k}1}): A_{i}\in {\cal B}_x  \right\}$. Then $C_x$ is an identity submatrix of $C$ of size $n$.
\end{lemma}
\begin{IEEEproof}
Note that there are $n$ rows in submatrix $C_x$ of $C$. Now we show that there are $n$ columns. By the property of TD with $\lambda =1$, the elements $\{x,A_{i}(G_{j_{i}{\oplus_k}1}) \}\subset A_{i}$ only and does not lie in any other $A_{i'}$ for any $i'\neq i$. Thus the elements $A_{i}(G_{j_{i}{\oplus_k}1})\neq A_{i'}(G_{j_{i'}{\oplus_k}1})$ for any $i'\neq i$. Hence, there are $n$ columns in $C_x$.

Fix any arbitrary row indexed by $A_{i} \in {\cal B}_x$. Suppose this row has entry $1$ in two distinct columns indexed by $y$ and ${y'}$ of submatrix $C_x$ where $y,{y'} \in {\cal X}$. This means that $y,{y'} \in A_{i}$ and thus $\{x,y,y'\} \subseteq A_i$, which means $\{x,y\} \nsubseteq A_i'$ and $\{x,y'\} \nsubseteq A_i'$ for any $i\neq i'$.	 This means both $y,{y'} \in G_{j_{i}{\oplus_k}1}$. By the property (TD4) for $\lambda=1$, this is not possible. Hence, each row of submatrix $C_x$ has entry $1$ in only one column.

Suppose the column $A_{i}(G_{j_{i}{\oplus_k}1})$ has $1$ in two rows, say $A_{i}$ and $A_{i'}$ such that $i \neq {i'}$. Then the set $\left\{x,A_{i}(G_{j_{i}{\oplus_k}1}) \right\}$ occurs in both blocks $A_{i}$ and $A_{i'}$. Thus we have a contradiction with the property of Transversal design for $\lambda=1$ i.e. the pair $\{x,A_{i}(G_{j_{i}{\oplus_k}1}) \}\in {\cal X}$ is contained in either one group or in exactly one block, but not both. Hence, $C_x$ is an identity submatrix of $C$ of size $n$.  
\end{IEEEproof}

\begin{lemma}
\label{TD_main}
For distinct $x_1,x_2\in{\cal X} $, there is no $x\in {\cal X}$, $A \in {\cal B}$ with $C(A,x)=1$ such that $C(A,x)$ is covered by both $C_{x_1}$ and $C_{x_2}$. Further, the set of matrices $\{C_x :x\in {\cal X}\}$ forms a non-overlapping identity submatrix cover of $C$ of size $n$.
\end{lemma}
\begin{IEEEproof}
The first part of the proof follows the same arguments as Lemma \ref{no_overlap}. The total number of $1$'s in matrix $C$ is equal to the product of number of $1$'s in each row and the number of rows, and thus equal to $kn^2$. For each $x\in {\cal X}$ (note that $|{\cal X}|=kn)$, there exists an identity submatrix $C_x$. From the first part of the proof and Lemma \ref{lemma:td_Cx_identity}, we see that each identity submatrix is of size $n$ with no overlaps. These identity submatrices will cover $kn^{2}$ number of $1$'s in matrix $C$ which is equal to total number of $1$'s in $C$. Hence, $\{C_x :x\in {\cal X}\}$ forms a non-overlapping identity submatrix cover of $C$ of size $n$. 
\end{IEEEproof}
We thus have the following theorem summarizing the caching scheme.
\begin{theorem}
\label{td_main_th}
The transpose of the incidence matrix of a TD-($k,n$) forms a $\left(K=n^{2},F=kn,(1-\frac{M}{N})=\frac{1}{n}\right)$ caching matrix. Further there is a transmission scheme with rate $R = 1$.
\end{theorem}

\begin{IEEEproof}
The parameters of the caching matrix $C$ have already been defined. By Lemma \ref{lemma:td_Cx_identity} and \ref{TD_main}, we have an identity submatrix cover consisting of $kn$ identity submatrices. Hence in Theorem \ref{rate def}, $S=kn$ and rate $R  =\frac{S}{F}=1$.
\end{IEEEproof}

\begin{example}
Consider the TD$(4,3)$ as given in example \ref{ex5}. We describe some identity submatrix $C_2$ corresponding to element `$2$' in ${\cal X}$. Element `$2$' is present in blocks $\{\{2,4,9,11\},\{2,5,7,12\},\{2,6,8,10\}\}$. Hence, the rows of the identity submatrix matrix $C_2$ are indexed by $\{2,4,9,11\}, \{2,5,7,12\}, \{2,6,8,10\}$ and columns are indexed by $4,5,6$ (since $4,5,6$ are the next elements present after `$2$' in blocks $\{2,4,9,11\}, \{2,5,7,12\}, \{2,6,8,10\}$ respectively) of matrix $C$. In this manner, the submatrices $C_i: i \in 
{\cal X}$, gives us a rate $1$ transmission scheme.
\end{example}



\section{$t$-subspace design based Coded Caching Scheme}
\label{sec:subspace_designs_scheme}

Let $({\cal V},{\cal A})$ denote a $t$-$(v, k, 1)_q$-subspace design for $t\geq 2$. Let the blocks in this design  be denoted by ${\cal A} = \{B_1, B_2, \dots, B_b\}$.
Let $T$ denote the set of all $1$-dim subspaces of ${\cal V}$, $H$ denote the set of all $t$-dim subspaces of ${\cal V}$, and $R$ denote the set of all $(t-1)$-dim subspaces of ${\cal V}$.

We construct a binary matrix $C$ as follows. Let the rows of $C$ be indexed by the set $R$. Let the columns be indexed by $\{(y,B): y \in T, y \subset B, B \in {\cal A}\}$. The number of rows in matrix $C$ is $\gbinom{v}{t-1}$. The number of columns in matrix $T$ is $b\gbinom{k}{1}=\frac{\gbinom{v}{t}\gbinom{k}{1}}{\gbinom{k}{t}}$. For some $D \in R$, the matrix $C = (C(D,(y,B)))$ is defined by the rule,
\begin{equation*}
\newline C(D,(y,B)) =
  \begin{cases}
   1, & \text{if}\ D \bigoplus y \in H, D \bigoplus y \subset B \\
    0, & \text{otherwise}. \\
  \end{cases}    
\end{equation*}

 \begin{remark}  
\label{rem:subspacedesigncachingmatrix:constantcolweight}
The number of $1$'s in each row of $T$ is $\gbinom{v-t+1}{1} \cdot q^{t-1}$. We see this by noting that the distinct $r$-dim subspaces of $\mathbb{F}_q^v$ intersecting a fixed $s$-dim subspace in some $l$-dim subspace is $q^{(r-l)(s-l)}\gbinom{v-s}{r-l} \gbinom{s}{l}$ (see \cite{hirschfeld1998projective} for a proof). To count the number of $1$'s in each row, since each $t$-dim subspace is included precisely in one block of the design, we only have to count the number of $t$-dim subspaces of $\fq^v$ intersecting a fixed $(t-1)$-dim subspace (identified by the row) in a $(t-1)$-dim subspace. This turns out to be precisely  $\gbinom{v-t+1}{1} \cdot q^{t-1}$. Similarly, we can obtain that the number of $1$'s in each column is $\gbinom{k-1}{t-1}\cdot q^{t-1}$ by the expression above.  Hence, the binary caching matrix $C$ defined above is also a constant row and column weight matrix. 
\end{remark}
 
Thus, we see that the matrix $C$ gives a $\bigg(\gbinom{v}{t-1},\frac{\gbinom{v}{t}\gbinom{k}{1}}{\gbinom{k}{t}},\frac{\gbinom{v-t+1}{1}\gbinom{k}{t}q^{t-1}}{\gbinom{k}{1}\gbinom{v}{t}}\bigg)$-caching matrix.  For some $y \in T$, let ${\cal B}_y$ be the set of blocks containing $y$. Now, 
\begin{equation*}
  |{\cal B}_y| = \lambda_1 = \frac{\gbinom{v-1}{t-1}}{\gbinom{k-1}{t-1}}.  
\end{equation*}

Denote ${\cal B}_y$ by ${\cal B}_y = \{B_1, B_2, \dots, B_{\lambda_1}\}$. For any $B_i$, denote by $D_i = \{D_{i,j}: j \in \left[\gbinom{k-1}{t-1} \cdot q^{t-1}\right]\}$  the set of all $(t-1)$-subspaces of $B_i$ that do not contain $y$. In the next lemma, we describe an identity submatrix of the matrix $C$. 


\begin{lemma}
\label{lemma:idmatrix_subspdesigns}
For some $j \in \left[\gbinom{k-1}{t-1} \cdot q^{t-1}\right]$ and $y \in T$, consider the submatrix $C_{y,j}$ of $C$ defined as follows: the rows of $C_{y,j}$ are indexed by $\{D_{i,j}: \forall i \in [\lambda_1]\}$ and the columns are indexed by $\{(y,B_i): B_i \in {\cal B}_y\}$. Then $C_{y,j}$ is an identity submatrix of $C$ of size $\lambda_1$.
\end{lemma}
\begin{IEEEproof}
It is clear that the number of rows and columns in $C_{y,j}$ is equal to $\lambda_1$. Thus, $C_{y,j}$ is a square matrix of size $\lambda_1$.

Now, consider a row of $C_{y,j}$ indexed by $D_{i,j}$ for some $i \in [\lambda_1].$ Suppose a column $(y,B)$ for some $B \in {\cal B}_y$ has a $1$ in the row indexed by $D_{i,j}$. This implies $D_{i,j} \bigoplus y \subset B$, but according to the rule by which the matrix $C$ is defined, there is only one $B$ such that $D_{i,j} \bigoplus y \subset B$, which is precisely $B = B_i$. Thus, each row in the submatrix $C_{y,j}$ contains only one $1$.

Now, consider a column of $C_{y,j}$ indexed by $(y,B_i)$ some $B_i \in {\cal B}_y$. Suppose for some row $D_{i',j}$ has a $1$ in the column indexed by $(y,B_i)$. Then it implies that $D_{i',j} \bigoplus y \subset B_i$, which further means that $D_{i,j} \bigoplus y \subset B_i \cap B_{i'}$. But because of the property of $t$-subspace design with $\lambda = 1$, we have that $i = i'$. Thus, each column in the submatrix $C_{y,j}$ contains only one $1$. This proves the lemma.
\end{IEEEproof}

In the next two lemmas, we will prove that there is no overlap between the identity sub-matrices in the collection $\{C_{y,j}: \forall y\in T,~ \forall j \in \left[\gbinom{k-1}{t-1} \cdot q^{t-1}\right]\}$ and that these identity sub-matrices will cover all the entries where $C(D,(y,B)) =1$ in matrix $C$. 

\begin{lemma}
\label{lemma_2}
Any $C(D,(y,B)) =1$ such that $y\in T, B\in{\cal A}, D\in R$ will be covered by exactly one identity submatrix of $C$ (as defined in Lemma \ref{lemma:idmatrix_subspdesigns}).
\end{lemma}
\begin{IEEEproof}
Let $C(D,(y,B)) =1$ be covered by an identity submatrix $C_{{y'},j}$ of $C$.  As $C(D,(y,B)) =1$, we have that $D \bigoplus y \subset B$. By definition of $C_{{y'},j}$ in Lemma \ref{lemma:idmatrix_subspdesigns}, we must first have $y = {y'}$. Further, it must be that $D \subset B_i \backslash y$ for some $B_i$ which contains $y$. Therefore, $D \bigoplus y \subset B_i$, which means $B_i = B$ (as $\lambda=1$). Hence, the unique transmission which covers $C(D,(y,B)) =1$ is $C_{y,j}$ where $j$ is such that $D=D_{i,j}$ is the unique $j^{th}$ set in $D_i$. This completes the proof.
\end{IEEEproof}

\begin{lemma}
\label{lemma_3}
The set of matrices $\{C_{y,j}: \forall y\in T,~ \forall j \in \left[\gbinom{k-1}{t-1} \cdot q^{t-1}\right]\}$ forms a non-overlapping identity submatrix cover of $C$.
\end{lemma}
\begin{IEEEproof}
The total number of $1$'s in $C$ is equal to the product of the number of $1$'s in each row and the number of rows, and thus equal to $\gbinom{v-t+1}{1}\gbinom{v}{t-1}q^{t-1}$.
The size of each identity submatrix is 
$\frac{\gbinom{v-1}{t-1}}{\gbinom{k-1}{t-1}}$. Since we have an identity submatrix $C_{y,j}$ for each $j\in \left[\gbinom{k-1}{t-1} q^{t-1}\right],~ y\in T$, the number of identity submatrices is $\gbinom{k-1}{t-1} \gbinom{v}{1}q^{t-1}$.  
Moreover, from Lemma \ref{lemma_2}, there are no overlaps between the identity submatrices. Hence, the total number of $1$'s covered by all identity submatrices 
\begin{equation*}
   {\gbinom{k-1}{t-1}\gbinom{v}{1}q^{t-1}}{\frac{\gbinom{v-1}{t-1}v}{\gbinom{k-1}{t-1}}}=q^{t-1}\gbinom{v-1}{t-1}\gbinom{v}{1}. 
\end{equation*}
It is easy to verify that the above quantity is equal to the total number of $1$'s in matrix $C$. Hence, $\{C_{y,j}: \forall y\in T,~ \forall j \in \left[\gbinom{k-1}{t-1} \cdot q^{t-1}\right]\}$ forms a non-overlapping identity submatrix cover of $C$.
\end{IEEEproof}

We thus have the below theorem summarizing the caching
scheme.

\begin{theorem}
    \label{cc_sd}
The matrix $C$ of a $t$-($v,k,1)_q$-subspace design forms a $\bigg(\gbinom{v}{t-1},\frac{\gbinom{v}{t}\gbinom{k}{1}}{\gbinom{k}{t}},\frac{\gbinom{v-t+1}{1}\gbinom{k}{t}q^{t-1}}{\gbinom{k}{1}\gbinom{v}{t}}\bigg)$-caching matrix. Further, there is a transmission scheme with rate $ R=\frac{\gbinom{k-1}{t-1} \gbinom{v}{1}\gbinom{k}{t}q^{t-1}}{\gbinom{v}{t}\gbinom{k}{1}}$.
\end{theorem}

\begin{example}
    We give an example for the $t=k$ scenario, which leads to a $k$-$(v,k,1)_q$ subspace design. Let $q=4$, $v=3$, and $k=t=2$. Then we get a coded caching scheme with parameters $K=21$, $1-M/N=0.19,$ $F=105$, and rate $R=0.8$. This is also one of the numerical examples compared in Table \ref{comparison_man} in Section \ref{sec:numerical}. 
\end{example}

\section{System Model for Coded MapReduce Distributed Computing}
\label{CDR:systemmodel}
In the forthcoming sections, we discuss the application of the  designs-based binary matrix we have developed in the framework of Coded MapReduce introduced in \cite{CMR,RDC}. Towards this end, we first briefly review the formal system model presented in \cite{RDC}, where the task is to compute $Q$ output functions on a large file using $K$ servers or computing nodes, which are indexed by a set ${\cal K}$. We denote $\beta\triangleq \frac{Q}{K}$, and assume $\beta$ is an integer as in \cite{RDC}. The file is divided into $F$ subfiles (we assume $F\geq K$), and denoted by a set ${\cal F}$ of size $F$ (abusing the notation a bit, we will also use $\cal F$ to index the $F$ subfiles of the file). The parameter $F$ is also referred as \textit{file complexity} in the context of coded distributed computing. Each subfile is assigned to $r$ servers where $r$ is called \textit{computational load}. Clearly, $r\geq 1$. Let ${\cal M}_k \subseteq {\cal F}$ denote the set of the subfiles assigned to server $k$, $k\in {\cal K}$.

Let the $Q$ output functions to be computed be denoted as $\phi_1,\hdots,\phi_Q$ where each $\phi_q$ maps all the input files to $u_q$, where $u_q = \phi_q(\{\forall f\in{\cal F}\})$ is a binary stream of some fixed length. Let $g_{q,f} , \forall q\in [Q], \forall f\in {\cal F}$ denote the \textit{map functions} which maps the input subfile $f\in {\cal F}$ into $Q$ intermediate values (IVAs), each a consisting of $T$ bits, denoted as $\{v_{1,f},v_{2,f}.\hdots,v_{Q,f} \}$. Each $v_{q,f}\triangleq  g_{q,f}(f), ~q\in [Q], ~f\in {\cal F} $ represents the IVA of length $T$ bits of the corresponding to the $q^{th}$ function and the subfile $f$. The \textit{reduce function} denoted by $h_q , q\in[Q]$ maps IVAs $v_{q,f}: \forall f\in {\cal F}$ to output bit stream $u_q$. Thus, we have $u_q = \phi_{q}(\{\forall f\in {\cal F}\}) = h_q(\{v_{q,f}: \forall f\in {\cal F}\}) = h_q(\{g_{q,f}(f): \forall f\in {\cal F}\})$. A distributed computing scheme in the MapReduce framework consists of three phases: map, shuffle and reduce phases, which we describe using the above functions as follows. 

\begin{enumerate}
\item \emph{Map Phase:} In map phase, each server $k\in {\cal K}$ will compute all the IVAs for the subfiles in ${\cal M}_k$ using the map functions, i.e., server $k$ computes $g_{q,f}(f): \forall f\in{\cal M}_k,\forall q$. Thus, after the map phase, server $k\in {\cal K}$ has $\{v_{q,f} : \forall q\in [Q], \forall f\in {\cal M}_k \}$. 
\item \emph{Shuffle Phase:} Each server is responsible for \emph{reducing} (computing the $h_q$ functions) a distinct subset of $\beta= \frac{Q}{K}$ functions of the $Q$ functions. Let ${\cal W}_{k}=\{q_{(k,b)}:\forall b\in[\beta]\}\subset [Q]$ denote the indices of the functions to be reduced at server $k\in {\cal K}$, where $\cup_{k=1}^K{\cal W}_k=[Q]$ . For a server to compute the output of a reduce function, it needs the IVAs of that output function for all the subfiles. Apart from the IVAs already computed in the map phase corresponding to the subfiles in ${\cal M}_k$, each server $k\in {\cal K}$ further requires $\{v_{q,f} : \forall q\in {\cal W}_{k} , \forall f\not\in {\cal M}_k \}$ to reduce the functions assigned to it. Hence, in the shuffle phase, the servers send broadcast transmissions to each other to make sure that each server receives the IVAs it needs for performing the reduce operations assigned to it. 
\item \emph{Reduce Phase:} 
With the received IVAs in the shuffle phase and the IVAs computed locally in the map phase, server $k$ uses the reduce functions $h_q$ to compute the task assigned to it, i.e., the node $k$ computes $h_q(\{v_{q,f}: \forall f\in {\cal F}\})$ for each $q\in{\cal W}_k$, thus computing the value of the functions $\phi_q: \forall q\in{\cal W}_k$ on the input file effectively. 
\end{enumerate}
 
As in \cite{RDC} the \textit{normalized communication load} $L$ of a distributed computing framework is defined as the (normalized) total number of bits communicated in shuffle phase by all the $K$ servers and can be calculated as 
\[
L \triangleq \small \frac{\text{Total number of bits transmitted in shuffle phase}}{\text{$QFT$}},
\]
where $T$ is the size of each IVA in bits. The coded distributed computing framework introduced in \cite{CMR}, uses coded transmissions in the shuffle phase to reduce the communication load. The communication load achieved by the scheme in \cite{CMR,RDC} for computational load $r$ is shown to be $\frac{1}{r}(1-\frac{r}{K})$ and this communication load is shown to be optimal in \cite{RDC}.

\section{Binary matrices and Distributed Computing}
\label{binarymatricesforcomputing}
In Section \ref{sec:binarymatrices_cc_model},  we used binary matrices to design coded caching schemes. In a similar vein, in this section we describe how a distributed computing scheme can be derived from a binary matrix with constant column weight. 
\begin{definition}
[Binary Computing Matrix] Consider a binary matrix $C$ with rows indexed by a $K$-sized set ${\cal K}$ and columns indexed by a $F$-sized set ${\cal F}$  such that the number of $0$'s in any column is constant (say $r$).
Then the matrix $C$ defines a distributed computing scheme with $K$ users (indexed by ${\cal K}$), file complexity $F$ (subfiles indexed by ${\cal F}$) and computation load $= r$ as follows:
\begin{itemize}
\item Server $k\in {\cal K}$ maps subfile $f: \forall f \in {\cal F}$ if $C(k,f) = 0$ and does not map it if $C(k,f) = 1$.
\end{itemize}
We then call the matrix $C$ as a $(K,F,r)$-computing matrix. 
\end{definition}

Now, we present an example for a computing matrix, for which later in this section we also illustrate our new shuffling scheme.
\begin{example}
\label{DC_matrix}
Consider a set system $({\cal K},{\cal F})$ given by ${\cal K} = \left\{1,2,3,4,5,6,7 \right\}$
and ${\cal F} = \{127,145,136,467,256,357, \newline 234 \}$.
The incidence matrix $C$ for this set system is
\renewcommand{\kbldelim}{(}
\renewcommand{\kbrdelim}{)}
\[
  \small \text{$C$} = \kbordermatrix{
    & 127 & 145 & 136 & 467 & 256 & 357 & 234 \\
    1 & 1 & 1 & 1 & 0 & 0 & 0 & 0\\
    2 & 1 & 0 & 0 & 0 & 1 & 0 & 1\\
    3 & 0 & 0 & 1 & 0 & 0 & 1 & 1\\
    4 & 0 & 1 & 0 & 1 & 0 & 0 & 1\\
    5 & 0 & 1 & 0 & 0 & 1 & 1 & 0\\
    6 & 0 & 0 & 1 & 1 & 1 & 0 & 0\\
    7 & 1 & 0 & 0 & 1 & 0 & 1 & 0
  }.
\]
It is easy to see that gives us a $(7,7,4)$-computing matrix. The corresponding distributed  computing system has $K=7$ nodes, $F=7$ subfiles, in which each subfile is stored in $r=4$ users. For instance, subfile indexed by $127$ is stored in users $3,4,5$ and $6$. 
\end{example}



In order to describe the shuffle phase in which we do coded transmissions, we first describe a single round of two transmissions based on an identity submatrix of the computing matrix, which will serve a number of servers. 



\begin{lemma}
\label{lemma_identitysubmatrix}
Consider an identity submatrix of $C$ given by rows $\{k_1,k_2,..,k_l: k_i \in {\cal K}\}$ and columns $\{f_1,f_2,..,f_l: f_i \in {\cal F}\}$, such that $C(k_i,f_i)=1, \forall i \in [l]$, while $C(k_i,f_j)=0, \forall i,j \in [l]$ where $i\neq j$. Then there exists two transmissions of length $\beta T= \frac{QT}{K}$ bits each, one coded and one uncoded,  done by any two different servers $k_{i}$ and $k_{j} : i,j\in [l], i\neq j$ such that each server $k_{i}: i\in [l]$  can recover the missing IVAs,  $\{v_{q_{(k_{i},b)},f_{i}} :  \forall b\in[\beta]\}$, from these two coded transmissions.
\end{lemma}
\begin{IEEEproof}
By the definition of identity submatrix, for each $i \in [l],$ the IVAs  $\{v_{q_{(k_{i},b)},f_{i}} : \forall b\in[\beta]\}$ are not available at server $k_i$ but are available at the other servers $\{k_1,k_2,\hdots,k_l\} \backslash k_i$.
Therefore, for some $p\in [l]$, consider the coded transmission of length $\beta T$ by the server $k_p$
\[
\left\{
\sum\limits_{\substack{i=1 \\ i\neq p}}^l {v_{q_{(k_{i},1)},f_{i}}},
\sum\limits_{\substack{i=1 \\ i\neq p}}^l {v_{q_{(k_{i},2)},f_{i}}}, \hdots,
\sum\limits_{\substack{i=1 \\ i\neq p}}^l {v_{q_{(k_{i},\beta)},f_{i}}}
\right\}.
\]
From the above transmission,  each $i \in [l] \backslash p$ can clearly recover the IVAs $\{v_{q_{(k_{i},b)},f_{i}} : \forall b\in [\beta]\}$. For instance, from transmission $\sum\limits_{\substack{i=1 \\ i\neq p}}^l {v_{q_{(k_{i},1)},f_{i}}}$, server $k_j: j\in [l]\backslash p$ can recover the intermediate value $v_{q_{(k_{j},1)},f_{j}}$ as all the other intermediate values ${v_{q_{(k_{i},1)},f_{i}}}: i\in [l]\backslash \{j,p\}$ are already present at server $k_j$ from map phase. Now, pick any server $k_i : i \in [l], i \neq p$. Let this server $k_i$ transmit the uncoded transmission of size $\beta T$ bits as follows.
\[
\{ v_{q_{(k_{p},1)},f_{p}} ,~ v_{q_{(k_{p},2)},f_{p}}, \hdots, v_{q_{(k_{p},\beta)},f_{p}}\}.
\]
Thus, server $k_p$ receives the IVAs $\{v_{q_{(k_{p},b)},f_{p}} : \forall b\in[\beta]\}$ as is. This completes the proof.
\end{IEEEproof}
\subsection{A new simple low complexity coded data shuffling algorithm}
\label{lowcomplexitydatashufflingsubsection}
Using Lemma \ref{lemma_identitysubmatrix}, we present the following theorem, which gives a new simple scheme for the shuffling phase. 
\begin{theorem}
\label{DC_maintheorem}
Consider a computing matrix $C$ of size $K\times F$ with a non-overlapping identity submatrix cover $\mathfrak{C}= \{C_1,C_2,..,C_S\}$ where the size of each identity submatrix is $g\geq 2$. Then, there exists a distributed computing scheme with $K$ nodes, attaining computation load $r$ and communication load $L={\frac{2}{g}}{\left( 1- \frac{r}{K}\right)}$, with file complexity $F$.
\end{theorem}
\begin{IEEEproof}
By Lemma \ref{lemma_identitysubmatrix}, corresponding to each identity submatrix $C_i$ in $\mathfrak{C}$, there are two transmissions which exchanges all the missing IVAs (corresponding to all functions to be reduced at the respective servers captured by the row indices of $C_i$) with respect to the subfiles corresponding to columns of the submatrix $C_i$ amongst the users indexed by the rows of $C_i$. Note that since $\mathfrak{C}$ is an identity submatrix cover, each missing IVA at any user will be part of some transmission corresponding to some identity submatrix in $\mathfrak{C}$. Thus, all the missing IVAs at all the users corresponding to all the functions to be reduced, are decoded. Therefore the reduce functions can also be successfully executed at the respective nodes. As $\mathfrak{C}$ contains $S$ identity submatrices, the total number of transmissions are $2S$.  Each transmission is of size $\beta T=QT/K$ bits. Hence the communication load is given as
$L=\frac{2QTS}{KQFT}=\frac{2S}{KF}$. Each identity submatrix is of size $g$, then the total number of $1$'s in each identity submatrix is $g$. There is no overlap between the identity submatrices, therefore the total number of $1$'s in computing matrix is $Sg$. Also, the number of ones in each column of computing matrix is $K-r$ as each subfile is stored in $r$ servers. Hence the total number of $1$'s in matrix $C$ is given by,
\begin{align}
\label{DC_eqn}
    Sg=F(K-r)
\end{align}

Using (\ref{DC_eqn}) we have, $L=\frac{2(K-r)F}{KgF}={\frac{2}{g}}{\left( 1- \frac{r}{K}\right)}.$

\end{IEEEproof}
We now give an example illustrating our new scheme, continuing from Example \ref{DC_matrix}, showing an identity submatrix cover for the computing matrix shown in that example. 

\begin{example}[Continuation of Example \ref{DC_matrix}] 
The identity submatrices of the matrix in Example \ref{DC_matrix} shown using the $7$ shapes clearly form an identity submatrix cover consisting of $7$ non-overlapping identity submatrices. 

\begin{figure}[h]
  \centering
\includegraphics[height=3cm, width=5.8cm]{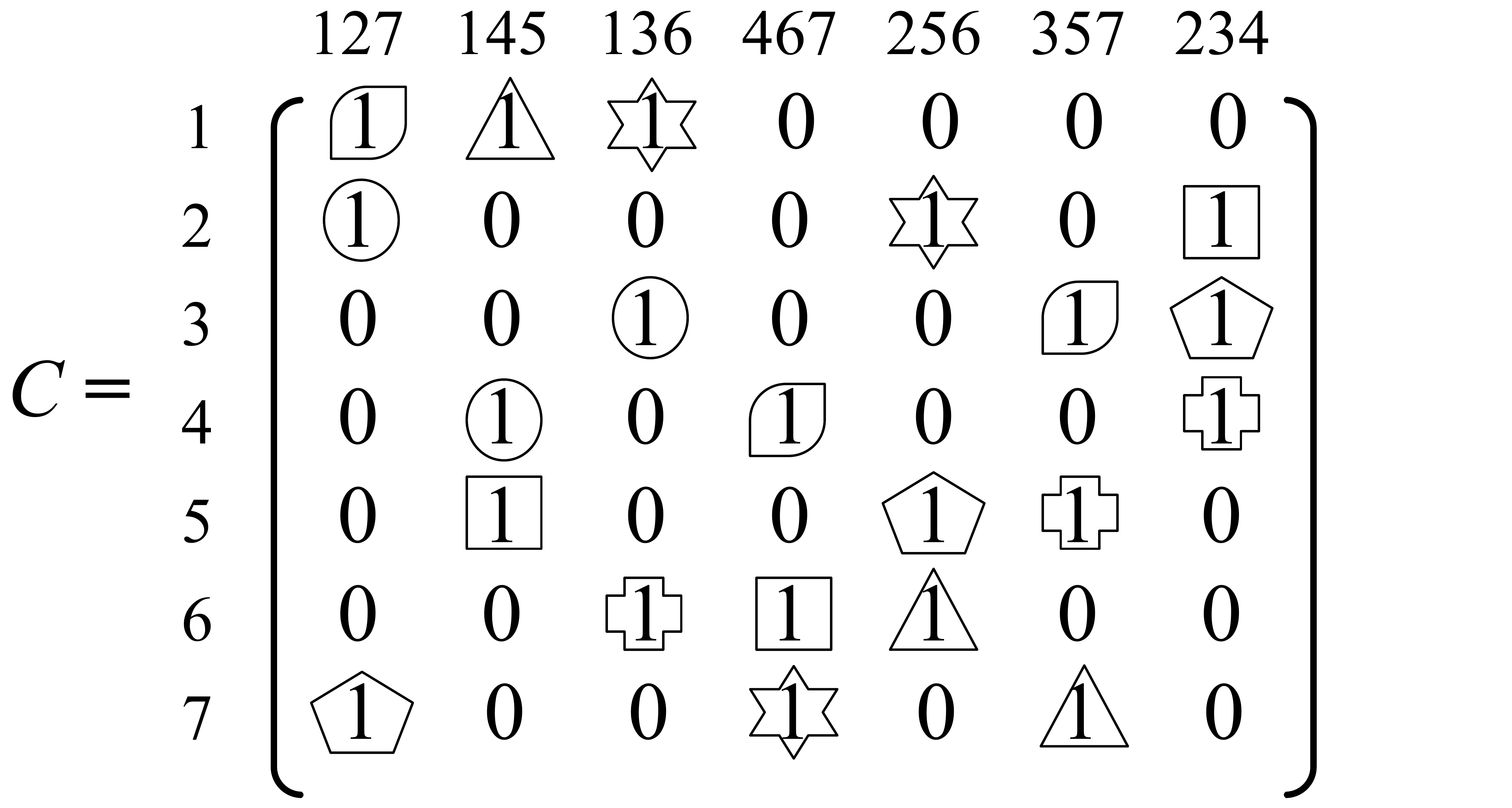}
\end{figure}
We now show one round of transmissions, consisting of two transmissions corresponding to one identity submatrix in the cover.  Let us consider one of the identity submatrix denoted as $C_2$, where $C_2$ is as below. \[
  \text{$C_2$} = \kbordermatrix{
    & 127 & 256  & 234 \\
    3 & 0 & 0 & 1 \\
    5 & 0 & 1 & 0 \\
    7 & 1 & 0 & 0 
  }
 \]
In $C_2$, one of the servers in $\{{3},{5},{7} \}$ will do the coded transmission and one will do the uncoded transmission. Let $Q=14$ (i.e., $\beta=2)$ and ${\cal W}_{3} =\{3,10\}, ~{\cal W}_{5}=\{5,12\},~{\cal W}_{7}=\{7,14\}$. Corresponding to this submatrix, the IVAs missing at server ${3}$ are $\{v_{3,234}, ~v_{10,234} \}$ as $234\notin{\cal M}_{{3}}$. Similarly, IVAs missing at server ${5}$ and ${7}$ are $\{ v_{5,256}, ~v_{12,256} \}$ and $\{ v_{7,127}, ~v_{14,127} \}$, respectively. Let server $3$ do the coded transmission and server $5$ do the uncoded transmission. Then the coded transmission by server $3$ is $\{ v_{7,127} \oplus v_{5,256} ,~ v_{14,127} \oplus v_{12,256} \}$. Observe that server $5$ already has $v_{7,127}$ and $v_{14,127}$ calculated in map phase, hence it can decode $v_{5,256}$ and $v_{12,256}$ from this coded transmission. Similarly, server $7$ can decode its required IVAs. The IVAs missing at server $3$ can be received by uncoded transmission $\{ v_{3,234}, ~v_{10,234}  \}$ done by server $5$. In a similar way, transmissions are done corresponding to each such identity submatrix in the identity submatrix cover, and decoding of the requisite IVAs by the respective servers can be done successfully, as per Theorem \ref{DC_maintheorem}. Observe that in this case the size of each identity submatrix is $3$, and thus the rate is $\frac{2}{3}(1-\frac{4}{7})=\frac{2}{7}$. 
\end{example}
The coded data shuffling scheme in \cite{RDC} achieves a communication load of $L=\frac{1}{r}(1-\frac{r}{K})$, and this was shown to be optimal. We now recall this scheme and present the map phase using a computing matrix, and calculate the communication load achieved by our new data shuffling scheme. This result is captured in the following corollary to Theorem \ref{DC_maintheorem}.

\begin{corollary}
\label{corrMaNschemeOurShuffling}
For any positive integers $K$ and $r\in[K]$, there exists a $(K,\binom{K}{r},r)$-computing matrix, from which we get a distributed computing scheme on $K$ nodes with computation load $r$ and communication load $L=\frac{2}{r+1}\left(1-\frac{r}{K}\right)$, with file complexity $F=\binom{K}{r}$. Further, this load $L<2L^*(r)$, where $L^*(r)$ is the optimal rate for a given computation load $r$.
\end{corollary}
\begin{IEEEproof}
Let ${\cal K}=[K]$ denote the set of nodes, and ${\cal F}=\{f_A:A\in\binom{[K]}{r}\}$ denote the set of subfiles, where $\binom{[K]}{r}$ denotes the $r$-sized subsets of $[K].$ Consider the matrix $C$ of size $K\times \binom{K}{r}$ with $C(k,f_A)=0$ if $k\in A$, and $C(k,f_A)=1$ if $k\notin A.$ The matrix $C$ is thus a $(K,F=\binom{K}{r},r)$-computing matrix. A non-overlapping identity submatrix cover of this matrix can be easily obtained as follows. Consider a subset of size $r+1$ of $[K],$ denoted by $B$. It is easy to check that the collection of rows defined by $B$ and the columns $B\backslash\{k\}:k\in B$ define an identity submatrix, which we denote by $C_B$. Further for each such $B$, the submatrix $C_B$ is of size $r+1$, and it is straightforward to check that these matrices are non-overlapping. Further, for $C(k,f_A)=1$ entry is covered by precisely that identity submatrix defined by $\{k\}\cup A.$ Thus, the collection of identity submatrices $\{C_B : B\in \binom{[K]}{r+1}\}$ is a non-overlapping identity submatrix cover of $C$, with $g=r+1$. Thus, by Theorem \ref{DC_maintheorem}, our data shuffling scheme on this computing matrix $C$ achieves a communication load $L=\frac{2}{r+1}(1-\frac{r}{K}).$ As  $L^*(r)=\frac{1}{r}(1-r/K)$ is known from \cite{RDC} to be the optimal communication load for computation load $r$, by comparing the two expressions we see that $L=\frac{2r}{r+1}L^*(r)<2L^*(r).$
\end{IEEEproof}
\subsubsection{Advantages of our shuffling scheme over the optimal scheme in \cite{RDC}}
\label{advantagessubsubsection}
Corollary \ref{corrMaNschemeOurShuffling} shows that our scheme has a higher communication load than the optimal scheme in \cite{RDC}. We now discuss some advantages of our scheme over the shuffling scheme in \cite{RDC}. During the data shuffling phase of the optimal-load scheme in \cite{RDC}, the IVAs have to be further subdivided into $r$ smaller chunks of length $\frac{T}{r}$ bits each. Then, for each subset $B$ of $[K]$ of size $r+1$, every server in $B$ encodes a set of $r$ chunks and broadcasts it to the other servers in $B$. The further chunking of the IVAs is absent in our scheme. This further dividing of the IVAs into $r$ smaller chunks incurs multiple costs including in coordination, indexing, switching, etc. which we now discuss. As a result of avoiding this IVA chunking, we refer to our scheme as a low complexity scheme compared to those in \cite{RDC}. 
\begin{itemize}
    \item Firstly, to identify the chunks, some indexing is required. This chunk-indexing cost is additional over and above the original file-complexity $F$. Our new scheme avoids this further chunking, and hence does not incur this cost. 
    \item Then, it is a requirement that multiple servers which have computed the same IVA in the map phase employ the exact IVA file chunking. If this is not done, decoding will not be possible. This decentralized IVA-chunking is thus unlike the original file complexity $F$, which is done prior to the placement in the storage of the nodes for mapping, and possibly in a single machine. This decentralized IVA-chunking of \cite{RDC} therefore requires some further coordination to establish agreement amongst the various nodes compared to our scheme.  
    \item Further, reading a large number of smaller sized chunks from the actual memory device (for instance, a hard disk or a flash drive) is more time and power consuming when compared to obtaining a smaller number of larger sized reads (our reads would be entire IVAs, i.e., $r$-times the size of the read in the scheme of \cite{RDC}). 
    \item Finally, suppose the transmissions at any node happen in a sequential manner following their occurrence in different sets $B\subseteq [K]$ of size $r+1$. Then since every node in $B$ participates in the transmission corresponding to $B$, this incurs the additional cost of turning the transmitting device ON and OFF a large number ($r\binom{K}{r}$) of times. However, in our scheme, only $2$ servers participate in the transmission round corresponding to any $B$. This means that we incur a cost of $2\binom{K}{r}$ number of switchings only. 
\end{itemize}
\subsection{Communication Load balancing of Scheme in Theorem \ref{DC_maintheorem}}
\label{loadbalancingsubsec}
The data shuffling scheme according to Theorem \ref{DC_maintheorem} ensures that only $2$ servers have to transmit for each identity submatrix in the identity submatrix cover $\mathfrak{C}$. As Lemma \ref{lemma_identitysubmatrix} chooses them arbitrarily, this may lead to a situation of imbalance in the communication load, i.e., some servers could be transmitting more bits while others transmit much less, or even don't transmit at all. This imbalance of network traffic may lead to other problems like node failures due to excess load, overall performance loss, etc. The following result shows that this problem of load imbalance can be rectified (provided some simple condition holds) by identifying two perfect matchings on an appropriately defined bipartite graph, which can be done in polynomial time in the parameter $S$. These conditions hold for some constructions we present, as well as some important schemes in literature, such as that in \cite{RDC}.
\begin{theorem}
\label{theorem_load balancing}
Let $C$ be a $(K,F,r)$-computing matrix with an non-overlapping identity cover $\mathfrak{C}=\{C_1,\hdots,C_S\},$ such that the size of each identity submatrix $C_i$ is $g\geq 2$ representing a computing system to reduce $Q$ functions. Then the coded data shuffling scheme in Theorem \ref{DC_maintheorem} is achievable with the property of load balancing, i.e. the total number of bits transmitted by each node is exactly $\frac{2S\beta T}{K}$ (where $\beta=Q/K$) out of which $\frac{S\beta T}{K}$ bits correspond to coded bits and the other $\frac{S\beta T}{K}$ bits correspond to uncoded bits, if (a) $\gamma\triangleq \frac{S}{K}$ is an integer, and (b) if each server $k$ appears in the row indices of the same number of identity submatrices in $\mathfrak{C}.$

\end{theorem}
\begin{IEEEproof} 
We first construct a bipartite graph ${\cal B}$. The set of left vertices of $\cal B$ are the set of servers (${\cal K}$) repeated $\gamma$ times, and is denoted as $\{k^{j}: k \in {\cal K}, j\in [\gamma]\}$. The right vertices are the indices of the identity submatrices in $\mathfrak{C}$ respectively. The edges are defined as follows. An edge between $k^j $ and $ C_{i}$ exists if and only if server $k$ is present in the row of identity submatrix $C_i$. Since the size of any identity submatrix in $\mathfrak{C}$ is $g$, the graph $\cal B$ is thus right regular with degree $g\gamma$, which means that the graph is biregular with degree $g\gamma$ as the cardinality of left vertices and right vertices are the same (namely, $\gamma K=S$), and by property (b).

A perfect matching on a bipartite graph is a matching $M$ (a collection of edges with no common vertices) such that every vertex in the graph is incident on at least one edge in the matching $M$. For regular bipartite graphs with $n$ vertices, a perfect matching can be found in time $O(n\log n)$ \cite{perfectM}. If such a perfect matching is found in $\cal B$, then for each $k\in{\cal K}$, each left vertex in the set $\{k^j:j\in[\gamma]\}$  is matched with precisely one right vertex, and thus the vertices $\{k^j:j\in[\gamma]\}$ are matched to right vertices (i.e., $\gamma$ identity submatrices), say $\{C_{k_i}:i\in[\gamma]\}$. In that case, we make the vertex $k$ responsible for the coded transmission corresponding to the identity submatrices $C_{k_i}:i\in[\gamma]$. Therefore each vertex $k$ is responsible for $\gamma$ coded transmissions. 

Now to define the server identity submatrix pairing for uncoded transmissions, we first obtain a new bipartite regular graph of degree $(g\gamma-\gamma)$ from $\cal B$ by removing some edges corresponding to the already-found perfect matching. Let for some $k,p$ and $i$, there exists an edge between $k^{p}$ and $C_{i}$ in perfect matching we have found, then we will remove all the edges between $k^{j}$ and $C_{i},~ \forall j\in [\gamma]$ from ${\cal B}$ . It is easy to see that the degree of each vertex of ${\cal B}$ is reduced to $g\gamma - \gamma=\gamma(g-1)$.  Note that as $g\geq 2$, we must have $\gamma(g-1)\geq 1$. Thus we have a new graph which is regular with degree $\gamma(g-1)$. Hence, we can find a perfect matching on this graph once again, using the algorithm in \cite{perfectM} for instance. By a similar argument as in the previous paragraph, for each $k\in{\cal K}$, we can get another set of $\gamma$ identities associated to it arising out of the new perfect matching say $\{C_{k_j}:j\in[\gamma]\}$. However, since the edges as described above are removed, we must have that $\{C_{k_i}:i\in[\gamma]\}\cap\{C_{k_j}:j\in[\gamma]\}=\phi$, i.e., none of the identities associated to $k$ in the first matching are repeated in the second. As a result, each $k\in{\cal K}$ can be assigned $\gamma$ uncoded transmissions corresponding to $\gamma$ identity submatrices which are all distinct from the $\gamma$ identities that $k$ has already been assigned to do coded transmissions for. 
Now, for each identity submatrix (right vertex), we have thus got two edges arising from the two perfect matchings obtained in the above manner, and these two must necessarily be incident on two nodes $k_1^{j_1},k_2^{j_2},$ where $k_1\neq k_2$. Thus, we have identified two distinct servers $k_1$ and $k_2$ from the first and second perfect matching doing the coded and uncoded transmission for each identity submatrix respectively. Combining the identification of these server nodes which are responsible for the coded and uncoded transmissions with the arguments of Theorem \ref{DC_maintheorem} which continue to hold as is, the proof is complete.
\end{IEEEproof}

\begin{figure}
  \centering
    \begin{subfigure}[b]{0.2\textwidth}
                \centering
                \includegraphics[height=4cm, width=3cm]{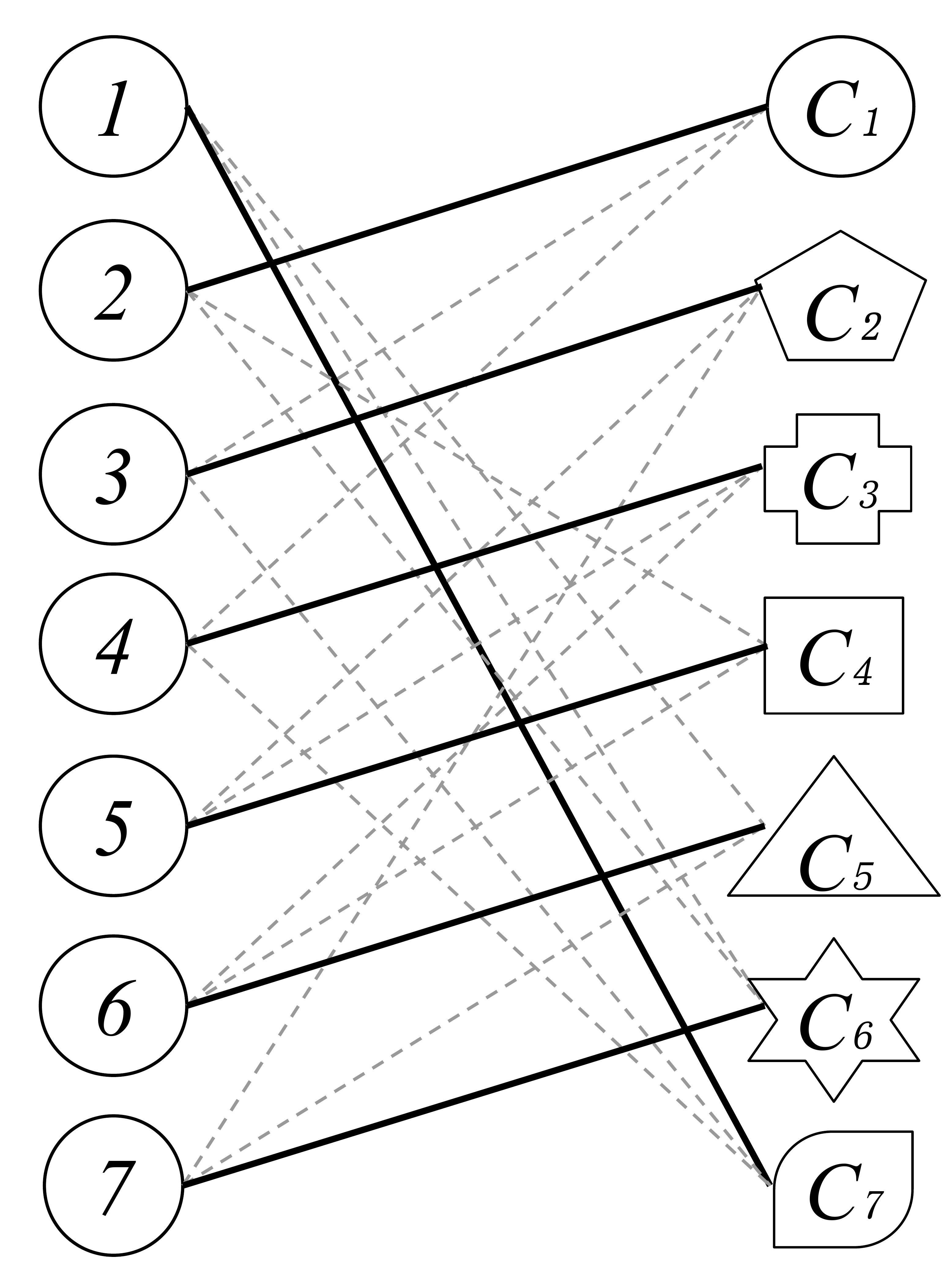}
        \end{subfigure}
        \begin{subfigure}[b]{0.19\textwidth}
                \centering
                \includegraphics[height=4cm, width=3cm]{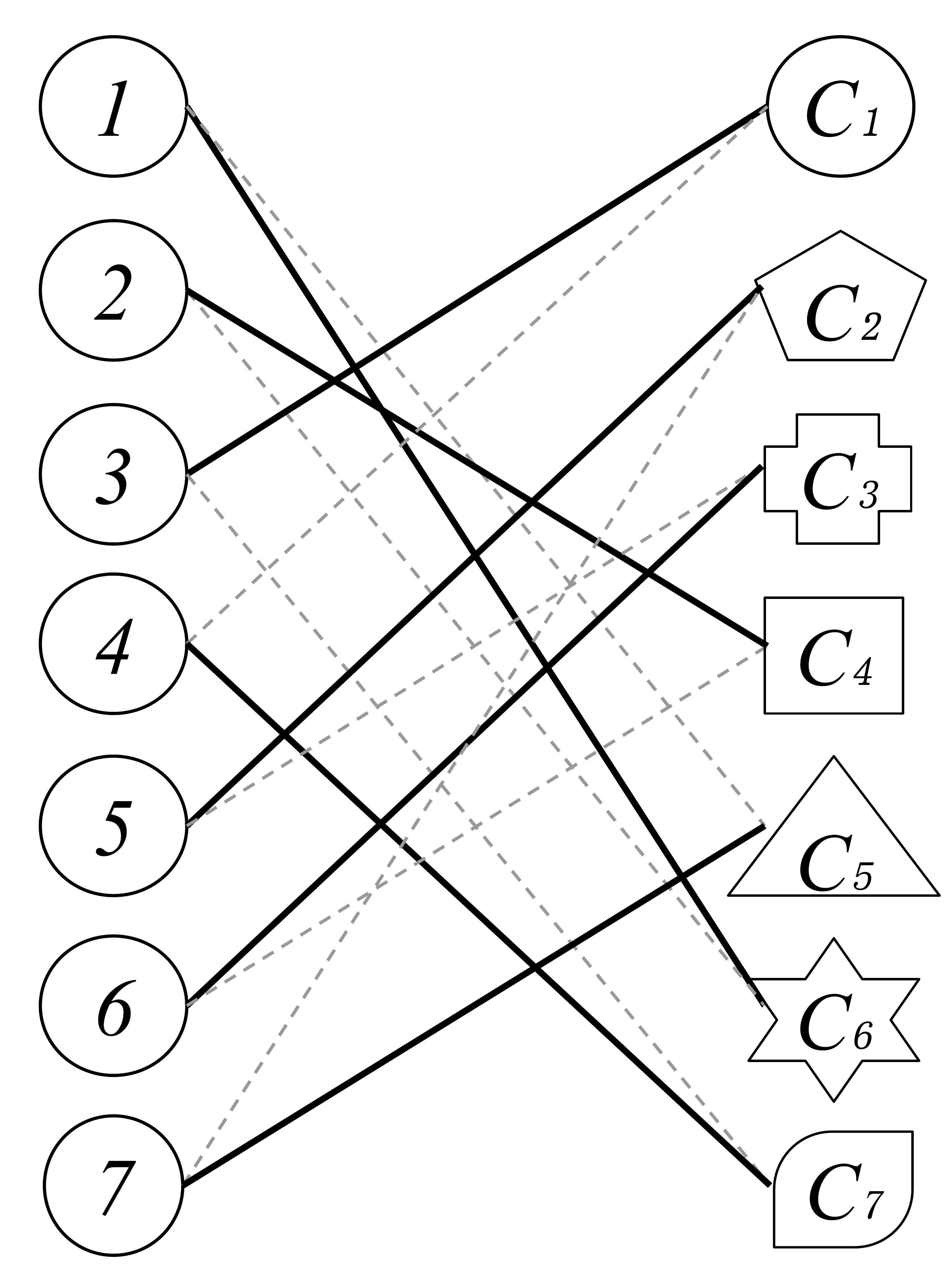}
        \end{subfigure}%
        \caption{Bipartite graphs based on the computing matrix in Example \ref{DC_matrix} illustrating Theorem \ref{theorem_load balancing}}
        \label{DC_graph}
        \hrule
       \end{figure}

\begin{example}
Let us consider the identity submatrix given in Example \ref{DC_matrix} for a distributed computing scenario where $K=7$ and $F=7$, for which an identity submatrix cover with $S=7$ matrices is shown in Example \ref{DC_matrix}. Thus $\gamma=\frac{S}{K}=1.$ For simplicity, we assume that $K=Q$ functions need to be reduced, thus $\beta=1$. The first figure on the left in Fig. \ref{DC_graph} shows the bipartite graph as constructed in Theorem \ref{theorem_load balancing}, with the users vertices on the left and the 7 identities on the right shown using the shapes. The bold edges in the first figure denote the first perfect matching obtained, using which coded transmissions are assigned. For instance, the user $6$ participates in the coded transmission with respect to the identity  submatrix corresponding to the shape `triangle' in Example \ref{DC_matrix}. After deleting these edges, we get the bipartite graph on the right, which is again a regular graph. The bold edges on this graph denote the perfect matching corresponding to the uncoded transmissions. Thus, each user is seen to participate in $2$ transmissions in this case, as $\beta=1,$ and $S=K=7$, transmitting $2T$ bits in total as given in Theorem \ref{theorem_load balancing}.

\end{example}

\begin{table*}
 \tabulinesep=1.0mm
\centering
\begin{tabular}{|M{2 cm}|c|c|c|c|c|}
 \hline
 
 \textbf{Combinatorial} & \textbf{Number of}   & \textbf{File } & \textbf{Computation Load} & \textbf{Communication}   &  \textbf{Communication}     \\

\textbf{ Design} & \textbf{servers} & \textbf{Complexity}  & $\textit{r}$ &  \textbf{ Load for} &  \textbf{ Load for $K-\kappa$}   \\

(${\lambda=1}$)  & $\textit{K}$ & $\textit{F}$ & & \textbf{non/partial straggler case} & \textbf{full straggler case}\\
 \hline

\textit{ BIBD}&$v$&$\frac{v(v-1)}{k(k-1)}$&$(v-k)$&$\frac{2k(k-1)}{v(v-1)}$& $\frac{2k(k-1)}{\kappa (v-1)}$ \\
\hline


 ${t}$-\textit{design } &$\binom{v}{t-1}$&$\frac{{\binom{v}{t}k}}{\binom{k}{t}}$&$\binom{v}{t-1}-\binom{k-1}{t-1}$& $\frac{2(v-t+1)\binom{k-1}{t-1}^2}{v\binom{v-1}{t-1}^2} $ & $\frac{2\binom{k-1}{t-1}^2}{\kappa \binom{v-1}{t-1}} $ \\
\hline


 \textit{ Transversal Design } &  $n^{2}$ & $kn$ &$n(n-1)$ & $\frac{2}{n^2}$ &  $\frac{2}{\kappa}$\\
\hline

 \textit{Subspace Design } & $\gbinom{v}{t-1}$ & $\frac{\gbinom{v}{t}\gbinom{k}{1}}{\gbinom{k}{t}}$ & $\gbinom{v}{t-1} - \gbinom{k-1}{t-1} q^{t-1}$ & $\frac{2\gbinom{k-1}{t-1}^2q^{t-1}}{\gbinom{v}{t-1}\gbinom{v-1}{t-1}}$ & $\frac{2\gbinom{k-1}{t-1}^2q^{t-1}}{\kappa \gbinom{v-1}{t-1}}$ \\
\hline
\end{tabular}
\caption{Parameters of distributed computing schemes based on combinatorial and subspace designs constructed in Sections \ref{sec:BIBD}-\ref{sec:subspace_designs_scheme}. Note that the parameters $v,k,t,n$ correspond to those of the respective designs.}
\label{DC_table}
\hrule
\end{table*}

\subsection{Low file complexity ($F$) schemes based on binary matrices from combinatorial designs}
\label{combinatorialdesigns}
In Sections \ref{sec:BIBD}-\ref{sec:subspace_designs_scheme}, combinatorial and subspace designs were used to construct constant row weight binary matrices which were used in coded caching. In Remarks \ref{rem:bibdconstweight}-\ref{rem:subspacedesigncachingmatrix:constantcolweight}, we see that each of these binary matrices also have constant column weight. Hence, these matrices also function as computing matrices. In Table \ref{DC_table}, we present the parameters of the distributed computing schemes derived from these designs-based computing matrices. Note that while the values of $K$ and $F$ in Table \ref{DC_table} (columns 2 and 3) correspond to the number of rows and columns respectively of the matrices designed in Sections \ref{sec:BIBD}-\ref{sec:subspace_designs_scheme}, the computation load $r$ is the difference between the number of rows and the column weight (obtained from Remarks \ref{rem:bibdconstweight}-\ref{rem:subspacedesigncachingmatrix:constantcolweight}) of these matrices. Further, the communication load (in column 5 of Table \ref{DC_table}) is calculated according to Theorem \ref{DC_maintheorem}, observing the size of the identity submatrices in the cover of each scheme, as obtained in Lemmas \ref{lemma:Cx_identity}, \ref{lemma:t5_1}, \ref{lemma:td_Cx_identity} and \ref{lemma:idmatrix_subspdesigns}. The last two columns of Table \ref{DC_table} pertains to the straggler scenario which we will consider in Section \ref{fullstragglersec}. In Section \ref{sec:numerical}, we compare these schemes with the baseline scheme from \cite{RDC} (which corresponds to the scheme in \cite{QYS} in the case of zero straggling nodes). Finally,  for load balancing according to Theorem \ref{theorem_load balancing}, we need the number of identity submatrices $S$ to be divisible by $K$ (number of servers). This  is satisfied only in the case of schemes from BIBDs and the transversal designs. The $t$-design based scheme never achieves the load balancing because this property $K$ does not divide $S$ in general, for this scheme. The same holds for the subspace design based scheme as well.

\section{Extensions to the Straggler Scenarios}
\label{fullstragglersec}
One of the practical challenges in the distributed computing framework is the presence of \textit{straggling nodes}. Straggling nodes are the nodes that perform operations slower than the other nodes. In this section, we utilize the advantage of our scheme (Theorem \ref{DC_maintheorem}, Lemma \ref{lemma_identitysubmatrix}) that only two servers are involved to communicate in each transmission round (corresponding to one identity submatrix). This advantage is used for straggler robustness upto a fixed number of stragglers, namely $g-2$, where $g$ is the size of any identity. In the \textit{full straggler} scenario considered in Section \ref{DC_FullStragglers}, the straggling nodes are slow to the extent they will not be able to complete any map task assigned to them, i.e. they can be considered as failed nodes. Thus, they are not involved in the map, shuffle or the reduce phase. A similar setting was assumed in \cite{QYS}, the scheme with which we shall compare. In the \textit{partial straggler} scenario which we consider in Section \ref{DC_partialStragglers}, a straggling node is slower than the other node by some slowdown factor, but they are not failed nodes. 

In the straggler scenario, our goal for the shuffling phase remains the same: to exchange messages between the nodes so that the IVAs for the reduce phase at the respective nodes are available. We thus redefine the communication load for the straggler scenario as 
\begin{align*}
L(\kappa)= \small \frac{\text{Number of bits transmitted in shuffle phase in `worst case'}}{\text{$QFT$}}
\end{align*}
where the `worst case' refers to the worst subset of $K-\kappa~ (or~ K-\kappa')$ stragglers (that subset which creates the largest load). 


\subsection{Full Stragglers}
\label{DC_FullStragglers}
In this section, we will discuss how to use the computing matrix to deal with full stragglers. The scheme we present here is robust upto $g-2$ stragglers. We consider a setup that as soon as $\kappa, \kappa\in{\cal K}$ servers complete the map operation, data shuffling phase can start.  We call this set of $\kappa $ servers to be \textit{surviving servers} while the rest $K-\kappa \in [0:g-2]$ servers are the full stragglers. For simplicity, we assume that $\frac{Q}{\kappa}$ is an integer, and thus $Q$ functions are evenly distributed among the $\kappa$ surviving servers. Below we describe a scheme which is robust for $g-2$ full stragglers. 
\begin{theorem}
\label{fullstragglertheorem}
Consider a \textit{computing matrix} of size $K\times F$ with a non-overlapping identity submatrix cover $\mathfrak{C}= \{C_1,C_2,..,C_S\}$ where the size of each identity submatrix is ${g\geq 2}$. Then, there exists a distributed computing scheme with $K$ nodes that is robust for $K-\kappa \in [0:g-2]$ full stragglers, attaining computation load $=r$ and communication load $L(\kappa)={\dfrac{2}{g}}{\left( \dfrac{K}{\kappa}- \dfrac{r}{\kappa}\right)},$ with file complexity $F$. 
\end{theorem}
\begin{IEEEproof}
Suppose some arbitrary $K-\kappa$ set of nodes failed. Since $K-\kappa\leq g-2,$ at least $2$ servers must still survive with respect to the rows of any identity submatrix $C_i$ in the cover $\mathfrak{C}.$ Thus, two servers for transmissions as in Lemma \ref{lemma_identitysubmatrix} are available for each identity submatrix. Hence a similar scheme as in Theorem \ref{DC_maintheorem} will be feasible, with the difference that since each server is allocated to reduce $\frac{Q}{\kappa}$ functions rather than $\frac{Q}{K}$ as in Theorem \ref{DC_maintheorem}. Following the similar arguments as Theorem \ref{DC_maintheorem}, the communication load in this case is calculated as follows.


\begin{center}
$\text{}~L =\dfrac{2QST}{\kappa QFT}=\dfrac{2S}{\kappa F}$ 
\end{center}
Using (\ref{DC_eqn}) we have,
\begin{center}
$L=\dfrac{2(K-r)F}{\kappa gF}={\dfrac{2}{g}}{\left( \dfrac{K}{\kappa}- \dfrac{r}{\kappa}\right)} \cdot$
\end{center}
Noting that we considered an arbitrary set of stragglers completes the proof.
\end{IEEEproof}

 





\begin{table*}
\centering
\begin{tabular}{|M{5 cm}|M{2.5 cm}|M{3.5 cm}|M{3 cm}|}
\hline
Distributed Computing Parameters of MAN-PDA & Number of non stragglers $\kappa$  & Optimal Communication Load in \cite{QYS} & Communication Load in Theorem \ref{fullstragglertheorem} \\
\hline
$K = 5, r=2,F=10,g=3$ & 5 & 0.3 & 0.4 \\
\hline
$K = 5, r=2,F=10,g=3$ & 4 & 0.45 & 0.5 \\
\hline
$K = 7, r=4,F=35,g=5$ & 7 & 0.107 & 0.171 \\
\hline
$K = 7, r=4,F=35,g=5$ & 6 & 0.13 & 0.2 \\
\hline
$K = 7, r=4,F=35,g=5$ & 5 & 0.17 & 0.24 \\
\hline
$K = 10, r=3,F=120,g=4$ & 10 & 0.23 & 0.35 \\
\hline
$K = 10, r=3,F=120,g=4$ & 9 & 0.27 & 0.39 \\
\hline
$K = 10, r=3,F=120,g=4$ & 8 & 0.3305 & 0.4375 \\
\hline
\end{tabular}

\caption{Numerical comparisons of communication load of scheme in \cite{QYS} and Theorem \ref{fullstragglertheorem}, for the computing matrix arising from the scheme in \cite{RDC} (Remark \ref{rem:fullstraggleroptimalschemecomparison}).}
\hrule
\label{DC_table2}
\end{table*}

\begin{remark} 
\label{rem:fullstraggleroptimalschemecomparison}In \cite{QYS} (and the extended version \cite{Q.Yan_PDA_CDC_stragglers}), the authors use the scheme of \cite{RDC} (as given in Corollary \ref{corrMaNschemeOurShuffling}) to achieve robustness against any set of $K-\kappa$ stragglers, where $K-\kappa \leq r-1$, and achieve a load given by the expression 
$$L^*(\kappa)=\left(1-\frac{r}{K}\right) \sum\limits_{\substack{i=r+\kappa-K}}^{min\{r,\kappa-1\}} {\frac{1}{i}\dfrac{\binom{r}{i}\binom{K-r-1}{\kappa-i-1}}{\binom{K-1}{\kappa-1}}}.$$ In fact, this load happens to be optimal (see \cite{Q.Yan_PDA_CDC_stragglers}) given parameters $K,r$, and number of stragglers $K-\kappa\leq r-1$. Following Corollary \ref{corrMaNschemeOurShuffling} and using Theorem \ref{fullstragglertheorem}, we see that the same scheme of \cite{RDC} which corresponds to $(K,F=\frac{K}{r},r)$-computing matrix (with $g=r+1$), we can obtain a distributed computing scheme that is robust against $K-\kappa$ stragglers (where $K-\kappa\leq g-2=r-1$), and has a load $L=\frac{2}{\kappa(r+1)}(K-r)$. We plug different values of $K,r,\kappa$ in the scheme of \cite{QYS} (which is shown to be optimal) and also for our scheme and compare communication loads for full straggler case in Table \ref{DC_table2}. We note that the load of our scheme is higher, though still comparable. Note that the file complexity is $F=\binom{K}{r}$ for both our scheme and that in \cite{QYS}. However, our scheme retains the advantages as mentioned in Section \ref{advantagessubsubsection}. 
\end{remark}

We further have tabulated in Table \ref{DC_table} (last column) the load obtained based from Theorem \ref{fullstragglertheorem} for the designs-based computing schemes obtained. in this work. Note that the load for the case of $(K-\kappa)$ full straggler nodes is $\frac{K}{\kappa}$ times the non-straggler load, as seen from Theorem \ref{fullstragglertheorem} and Theorem \ref{DC_maintheorem}. Numerical simulations comparing the communication loads of these schemes against the scheme from \cite{QYS} are presented in Section \ref{sec:numerical}. 

%

\subsection{Partial Stragglers}
\label{DC_partialStragglers}
In this section, we use the computing matrix to describe a distributed computing scheme with partial stragglers. Partial stragglers are the servers that perform the task assigned to them slower than the other servers, but they are not failed nodes. Partial stragglers take more time to complete the map phase as compared to other non straggling servers. We consider a scenario having $K$ servers with $\kappa'$ non-stragglers and  $K-\kappa' \in [0:g-2]$ partial stragglers. We give a scheme in which partial stragglers do not need transmit anything in the shuffling phase, however they remain responsible for reducing the output functions assigned to them. For a similar partial straggler setting, a distributed computing scheme was presented in \cite{vinayakPDA}.

We now briefly describe the scheme discussed in \cite{vinayakPDA}. The scheme in \cite{vinayakPDA}, uses the combinatorial structure known as $g$-regular PDA, which is very similar to the computing matrix with each identity submatrix of size $g$. Scheme in \cite{vinayakPDA} computes $Q=K$ output functions (i.e., $\beta=1$) and can handle $K-\kappa'<g-1$ partial stragglers. Also, in the map phase \cite{vinayakPDA}, all the servers including partial stragglers compute all the IVAs for the files stored in them i.e., server $k\in {\cal K}$ will compute $\{ v_{q_{(k,1)},f}: \forall q_{(k,1)}\in [Q], \forall f \in {\cal M}_{k} \}$ in the map phase. The communication load of the scheme in \cite{vinayakPDA} is given by $L=\frac{1}{g-1-(K-\kappa')}\left(1-\frac{r}{K}\right)$. 

In the present work, we propose a new scheme based on computing matrices that can handle partial stragglers. Our scheme can compute $Q=\beta K$ output functions ( where $\beta$ is an integer), while also requiring that the partial stragglers compute less number of IVAs as compared to in \cite{vinayakPDA}. For simplicity we assume that, $Q$ reduce functions are evenly distributed among $K$ servers. We divide the map phase in to three sub-phases. 
In our scheme, partial stragglers are involved only in map sub-phases 1 and 2 and in the reduce phase, but not in map sub-phase 3 or the shuffling phase. 


Let $C$ be a computing matrix of size $K\times F$  with a non-overlapping identity submatrix cover $\mathfrak{C}= \{C_1,C_2,..,C_S\}$ where the size of each identity submatrix is $g\geq 2$. Let ${\cal I}_k$ be the set of identity submatrices  which contain the server $k\in{\cal K}$ in its row index. As each $1$ in the $k^{th}$ row (for any $k\in {\cal K}$) of computing matrix $C$ corresponds to a different identity submatrix, hence, $|{\cal I}_k|$ is equal to the number of $1$'s in $k^{th}$ row of computing matrix $C$. For an identity submatrix $C'\in {\mathfrak{C}}$ of size $g$, we denote $\mathfrak{I}' = \{(k^{1}, f^{1}), \hdots,(k^{g}, f^{g}) \}$ be the set of indices from ${\cal K}\times {\cal F}$ where $1$ is present in submatrix $C'$. 


We divide the map phase into three sub-phases. Algorithm \ref{algo} gives the details of the IVA's calculated by each server in different map sub-phases. The servers calculate the IVAs in a \textit{sequence} as follows:
\begin{itemize}
    \item Map sub-phase 1: The servers will only calculate the IVAs for the functions they are reducing.
    
    \item Map sub-phase 2: Servers will calculate the IVAs (for functions which they need not reduce) which they require to decode the coded transmission from the identity submatrices (of the computing matrix) that they are involved in. 
    
    \item Map sub-phase 3: The first $\kappa'$ servers, which finishes map sub-phase 1 and 2, enters map sub-phase 3. We call this set of $\kappa'$ servers as an active set ${\mathfrak{D}}$. Once an active set ${\mathfrak{D}}$ comes into existence, the other $K-\kappa'$ servers (not in ${\mathfrak{D}}$) stop the mapping operation after completing map sub-phase 2. In map sub-phase 3, each server in active set ${\mathfrak{D}} $ will compute the IVAs for all its stored subfiles, which were not already calculated in map sub-phase 1 and 2.
    
\end{itemize}
Algorithm \ref{algo} presents this in a more formal way. Note that each server involved in $C'$ will compute $(g-1) \beta $ intermediate values. As each server appears in $|{\cal I}_k|$ identity submatrices, hence the partially straggling server calculates only $(g-1) \beta |{\cal I}_k|+\beta|{\cal M}_k|$ intermediate values, which is strictly smaller than $Q|{\cal M}_k|$ if $\beta>1$. After the completion of the map-phase as per Algorithm \ref{algo}, the IVAs are exchanged via coded transmissions made by the servers in the active set only. The following theorem describes the data shuffling phase of the computing scheme with map phase as described in Algorithm \ref{algo}, which is robust up to $g-2$ partial stragglers, where $g$ is a parameter of a computing matrix.

 \begin{algorithm}
\caption{Calculation of IVA's in Map sub-phases}
\label{algo}
\SetAlgoLined
\DontPrintSemicolon
\SetKwBlock{DoParallel}{do in parallel}{end}

\textit{Map sub-phase 1:}

\DoParallel
{
    for each $k\in {\cal K}$ calculate $ v_{q_{(k,b)},f} : \forall b\in[\beta], \forall f\in {\cal M}_k.$
}
\textit{Map sub-phase 2:}

\DoParallel
{
    for each $k\in {\cal K}$ \;
   \For{ $C' \in {\cal I}_k $}
   {
   calculate $v_{{q_{(k',b)}},f'} : \forall b\in [\beta], \forall (k', f') \in {\mathfrak{I}'}\backslash (k,f)$.
   }
}
Let ${\mathfrak D}$ denote the first $\kappa'$ servers to finish sub-phase 1 and 2.\;
\textit{Map sub-phase 3:}

\DoParallel
{
    for each $k\in {\mathfrak D}$
    calculate $v_{q_{(k,b)},f} : \forall q_{(k,b)} \in [Q], \forall f\in {\cal M}_k $, if not already calculated in sub-phase 1 and sub-phase 2.
}

\end{algorithm}

\begin{theorem}
\label{DC_partial_str}
Consider a computing matrix of size $K\times F$ with a non-overlapping identity submatrix cover $\mathfrak{C}= \{C_1,C_2,..,C_S\}$ where the size of each identity submatrix is $g\geq 2$. Then, there exists a distributed computing scheme with $K$ nodes that is robust for $K-\kappa'\in [0:g-2]$ partial stragglers, attaining computation load $=r$ and communication load $L(\kappa')={\dfrac{2}{g}}{\left( 1- \dfrac{r}{K}\right)} $, with file complexity $F$.
\end{theorem}

\begin{IEEEproof}
Suppose some arbitrary $K-\kappa'$ set of nodes straggle. Since $K-\kappa'\leq g-2,$ at least $2$ active servers must still survive with respect to the rows of any identity submatrix $C_i$ in the cover $\mathfrak{C}.$ Thus, two active servers for doing the transmissions as in Lemma \ref{lemma_identitysubmatrix} are available for each identity submatrix. 

Now the servers (both stragglers and non stragglers), in map sub-phase 2 of Algorithm \ref{algo}, has already calculated the IVAs which they require to decode the coded transmission. Hence, the servers in each identity submatrix will be able to decode the coded transmission, in order to get the missing (unmapped) IVAs, which they require to perform the reduce operation in reduce phase. Hence, a similar scheme as in Theorem \ref{DC_maintheorem} will be feasible, and the similar arguments can be followed as in Theorem \ref{DC_maintheorem} to derive the communication load.
\end{IEEEproof}

\section{Discussion}
\label{discussion}
     We have presented new coded caching and distributed computing schemes using binary matrices arising out of combinatorial designs and their $q$-analogs. Our schemes are primarily useful for the large local cache scenario. Taking wider classes of designs (especially those with higher values of the $\lambda$ parameter of the respective designs) are likely to result in lower cache size requirement. However, designing delivery schemes (identity submatrix covers) for these schemes appears difficult. Numerical results are shown with respect to some baseline schemes to illustrate the advantage of our schemes. Prior to earlier versions of this work \cite{Shailja_ISIT2020,CDesigns_Shailja_ISIT2019}, the work \cite{TaR} used special combinatorial structures called resolvable designs in a different way to construct coded caching schemes. After the appearance of \cite{Shailja_ISIT2020,CDesigns_Shailja_ISIT2019}, a few other works (for instance, \cite{Jian_PDA_CombiDesigns_BIBD_havetolook}) have taken our work forward as well. Special resolvable designs known as cross-resolvable designs were also used for the setting of multi-access coded caching (in which each client can connect to multiple caches) (for instance, see \cite{Katyal_CRD_multiaccess}).  We believe that the setting of cache-aided communications provides a fertile ground for further intensive application of the theory of combinatorial and subspace designs.

     In the case of MapReduce-based distributed computing, we have presented a new simple coded shuffling scheme which avoids further IVA-chunking as compared to the existing optimal scheme, at the cost of marginal increase in the communication load. We also presented conditions in which our new scheme achieves communication load balancing across the servers. We have also extended our schemes to the full and partial straggler scenarios. The presented distributed computing schemes using binary matrices arising from some combinatorial and subspace designs have the advantage of very small file-complexity schemes when compared to the optimal scheme for similar values of $K,r$, at the cost of having a higher rate. 

\section{Numerical Comparisons}
\label{sec:numerical}
In this section, we will provide numerical comparisons of our schemes with some existing important baseline schemes. While many other schemes have been proposed in the literature since the appearance of these baseline schemes, we primarily select these schemes for comparison because of their rate-optimality, or low-subpacketization requirement when achieving near-optimal rates. Because of these reasons, these schemes continue to be important and competitive in terms of their performance. 

With respect to coded caching, we present Tables \ref{comparison_man} and  \ref{comparison_sec}. Table \ref{comparison_man} presents some numerical examples of the coded caching schemes from Table \ref{tab1}. These are compared with the optimal-rate scheme from \cite{MaN} in terms of both subpacketization and rate, by equivalizing the number of clients $K$ and the uncached-fraction $1-\frac{M}{N}$. In most cases, we observe that the subpacketization $F$ of our schemes is lesser than the subpacketization $F^*=\binom{K}{MK/N}$ of \cite{MaN} (in some cases, this difference is quite large), while the rate $R$ of our scheme is in general larger than the rate of $R^*=K(1-M/N)/(1+MK/N)$ of the scheme in \cite{MaN} (however, only by less than an order of magnitude). In Table \ref{comparison_sec}, we numerically compare our schemes with the scheme from \cite{strongedgecoloringofbipartitegraphs}. The scheme from \cite{strongedgecoloringofbipartitegraphs} is one of the best known schemes for the coded caching setup of \cite{MaN}, as it has offers better subpacketization levels than \cite{MaN} in general, without much increase in rate. The scheme in \cite{strongedgecoloringofbipartitegraphs} depends on the strong-edge coloring of a related bipartite graph, constructed using a set-theoretic method using four non-negative integer parameters, $m,a,b,\lambda$, where $a,b\leq m$ and $\lambda\leq \min(a,b).$ We denote the coded caching parameters of the scheme from \cite{strongedgecoloringofbipartitegraphs} as $K_{SEC}, (1-\frac{M}{N})_{SEC}, R_{SEC},$ and $F_{SEC}$, whose expressions are described in the caption of  Table \ref{comparison_sec}. In this case, it is somewhat difficult to exactly match our parameters with \cite{strongedgecoloringofbipartitegraphs}. We try to match the number of clients as much as possible and compare the rest. We see that in some cases, our scheme has advantages, while in others, the scheme from \cite{strongedgecoloringofbipartitegraphs} dominates.  

In Table \ref{comparison_straggler}, we provide some numerical examples of our designs-based constructions for coded computing schemes. We provide the parameters of the designs used and the resulting parameters of the coded computing scheme, the number of servers $K$, the file complexity $F$, the computation load $r$, along with the communication load for the no-straggler scenario as given by Theorem \ref{DC_maintheorem} (which matches the partial straggler load as given by Theorem \ref{DC_partial_str}), and the full-straggler load (as given by Theorem \ref{fullstragglertheorem}) for $K-\kappa\in\{1,2\}$ stragglers. Finally, in Table \ref{comparison_qys}, we compare our schemes with that from \cite{QYS}. We see that our schemes have advantages in file-complexity, with increased communication loads. 

\begin{table*}[ht]
\centering
\begin{tabular}{|c|c|c|c|c|c|c|c|}
\hline
Combinatorial Design & Parameters (from specific constructions & $1-\frac{M}{N}$ &$K$ & $R$ & $F$ & $R^{*}$ from \cite{MaN}  & $F^{*}$ from \cite{MaN} \\ 
 $(\lambda=1)$ &   of the design) & &  & & &  &  \\
\hline
\textit{Symmetric BIBD,}  & $n=5$  & 0.193  & 31 & 1 & 31  & 0.23 & $7.36 \times 10^5$\\ 
Section \ref{sc_bibd} & & & & & & & \\
\hline
\textit{BIBD,}  & $n=5$  & 0.2  & 25 & 0.83 & 30  & 0.24 & $5.3 \times 10^4$\\ 
Section \ref{sc_bibd} & & & & & & & \\
\hline
\textit{$t$-design}  & $q=2$ & 0.1 & 10 & 0.16 & 15 & 0.1 & 10\\ 
Section \ref{sc_steiner} & & & & & & & \\
\hline
\textit{Tranversal Design,}  & $q=5$  & 0.2  & 25 & 1 & 25  & 0.24 & $5.3 \times 10^4$\\ 
Section \ref{sc_tranversal} & & & & & & & \\
\hline
\textit{\textit{Subspace Design,}}  & $2-(3,2,1)_2$ & $0.28$ & $7$ & $0.67$ & $21$ & $0.324$ & $21$ \\ 
{Section \ref{sc_subspace}} & & & & & & & \\
\hline
\textit{Subspace Design,}   & $2-(4,2,1)_2$ & $0.133$& $15$ & $0.28$ & $105$ & $0.142$ & $105$\\ 
{Section \ref{sc_subspace}} & & & & & & & \\
\hline
\textit{Subspace Design,}  & $2-(3,2,1)_3$ & $0.23$ & $13$ &  $0.75$ & $52$  & $0.27$ & $286$ \\ 
{Section \ref{sc_subspace}} & & & & & & & \\
\hline
\textit{Subspace Design,}    & $2-(3,2,1)_4$ & 0.19& 21& 0.8&  105& 0.22 & 5985\\ 
{Section \ref{sc_subspace}} & & & & & & & \\
\hline
\textit{Subspace Design,}   & $2-(3,2,1)_5$ & $0.16$ & $31$ & $0.83$ & $186$ & $0.183$ & $169911$ \\ 
{Section \ref{sc_subspace}} & & & & & & & \\
\hline
\textit{Subspace Design,} 
& $2-(3,2,1)_7$ & $0.122$ & $57$ & $0.875$ & $456$ & $0.136$ & $2.6 \times 10^8$ \\ 
{Section \ref{sc_subspace}} & & & & & & & \\
\hline
\end{tabular}
\caption{Numerical comparison of Coded Caching Schemes based on specific constructions of this paper, with the lower bounds of \cite{MaN}. Note that these schemes actually exist, as the designs are known to exist.}
\label{comparison_man}
\hrule
\end{table*}

\begin{table*}[ht]
\centering
\begin{tabular}{|M{2 cm}|c|c|c|c|c|c|c|c|c|}
\hline
Combinatorial & Parameters & $1-\frac{M}{N}$ &$K$ & $R$ & $F$ & $(1-\frac{M}{N})_{SEC}$ &$K_{SEC}$ & $R_{SEC}$ & $F_{SEC}$\\
Design & (from specific constructions & & & & & & & &  \\
($\lambda=1$) & of the design and SEC scheme) & & & & & & & & \\
\hline

\textit{BIBD} & $n=4$, $(m,a,b,\lambda)=(6,2,3,0)$ & 0.25 & 16 & 0.8 & 20 & 0.2 & 15 & 0.3 & 20\\
\hline

\textit{BIBD} & $n=7$, $(m,a,b,\lambda)=(8,4,3,3)$  & 0.143 & 49 & 0.875 & 56 & 0.071 & 70 & 0.143 & 56\\
\hline

\textit{Symmetric BIBD} & $n=4$, $(m,a,b,\lambda)=(7,2,2,0)$ & 0.238 & 21 & 1 & 21 & 0.476 & 21 & 1.667 & 21\\
\hline

\textit{Symmetric BIBD} & $n=7$, $(m,a,b,\lambda)=(8,3,3,3)$ & 0.14 & 57 & 1 & 57 & 0.018 & 56 & 0.018 & 56 \\
\hline

\textit{Tranversal Design} & $q=9$, $(m,a,b,\lambda)=(9,3,3,2)$ & 0.111 & 81 & 1 & 81 & 0.214 & 84 & 0.857 & 84\\ 
\hline

\end{tabular}
\caption{Numerical comparison of Coded Caching Schemes based on specific constructions of Combinatorial Designs (Section \ref{sec:summary_codedcaching}) with Strong Edge Coloring (SEC) Scheme\cite{strongedgecoloringofbipartitegraphs}. Note that $(1-\frac{M}{N})_{SEC}=\frac{\binom{a}{\lambda}\binom{m-a}{b-\lambda}}{\binom{m}{b}})$ , $K_{SEC}=\binom{m}{a}$, $R_{SEC}=\frac{\binom{m}{a}\binom{a}{\lambda}\binom{m-a}{b-\lambda}}{\binom{m}{b}\max\left\{\binom{a+b-2\lambda}{a-\lambda},\binom{m-a-b+2\lambda}{\lambda}\right\}}$, $F_{SEC}=\binom{m}{b}$.} 
\label{comparison_sec}
\end{table*}

\begin{table*}[ht]
\centering
\begin{tabular}{|M{2.5 cm}|c|c|c|c|c|c|c|}
\hline
Combinatorial & Parameters & Number of  & File   & Computation   & $L$ for  & $L$ for   & $L$ for \\
Design & (from specific  & servers & Complexity & Load  & non/partial & $K-\kappa=1$ & $K-\kappa=2$ \\
($\lambda=1$) & constructions & $K$ &$F$  & $r$&straggler case & & \\
& of the design) & & & & & & \\
\hline

\textit{BIBD}  & $n=7$ & 49 & 56 & 42 & 0.0357 & 0.0364 & 0.0372 \\
\hline

\textit{Symmetric BIBD } & $n=5$ & 31 & 186 & 26 & 0.0645 & 0.0667 & 0.0689 \\
\hline

\textit{t-design} & $q=3$ & 45 & 120 & 42 & 0.0111 & 0.0113 & 0.0116 \\
\hline


\textit{Transversal Design} & $q=5$& 25 & 25 & 20 & 0.08 & 0.083 & 0.087 \\
\hline

\textit{Subspace Design}  &$2-(3,2,1)_2$ & 7 & 21 & 5 & 0.19 & 0.22 & 0.27 \\
\hline
\textit{Subspace Design}  &$4-(5,4,1)_2$ & 155 & 465 & 147 & 0.00688 & 0.00693 & 0.00697 \\
\hline
\textit{Subspace Design}  &$3-(4,3,1)_3$ & 130 & 520 & 121 & 0.01065 & 0.01073 & 0.01082
 \\
\hline
\textit{Subspace Design}  &$4-(5,4,1)_3$ & 1210 & 4840 & 1183 & 0.0011157 & 0.0011166 & 0.0011175
  \\
\hline


\end{tabular}
\caption{Numerical comparisons of communication loads of our schemes  in non/partial straggler case and straggler case (applying the expressions in Table \ref{DC_table} to specific constructions of Section \ref{sec:summary_codedcaching}).}
\label{comparison_straggler}
\hrule
\end{table*}

\begin{table*}[ht]
\centering
\begin{tabular}{|M {2.3 cm}|c|c|c|c|c|c|c|}
\hline

Combinatorial  & Number of & Computation   & File  & File  & Number of   & Communication  & Optimal \\
Design & servers  & Load  & Complexity  & Complexity & non stragglers & Load in & Communication  \\
($\lambda=1$) & $K$ & $r$ & $F$ & $F$ in \cite{QYS} & $\kappa$ & Theorem \ref{fullstragglertheorem} & Load in \cite{QYS}  \\
\hline


\textit{BIBD } & 25 & 20 & 30 & 53130 & 25 & 0.067 & 0.01\\
\hline

\textit{BIBD } & 25 & 20 & 30 & 53130 & 23 & 0.072  & 0.011 \\
\hline

\textit{BIBD } & 121 & 110 & 132 & $1.28 \times 10^{15}$ & 119 & 0.0154 & 0.0008  \\
\hline



\textit{Symmetric BIBD } & 31 & 26 & 186 & 169911 & 31 & 0.065 &  0.006 \\
\hline

\textit{Symmetric BIBD } & 31 & 26 & 186 & 169911 & 29 & 0.069 &  0.0067 \\
\hline

\textit{Symmetric BIBD } & 133 & 122 & 1596 & $3.78 \times 10^{15}$ & 131 & 0.01526 & 0.00069 \\
\hline

\textit{$t$-design  } & 45 & 42 & 120 & 14190 & 45 & 0.0111 & 0.0016 \\
\hline

\textit{$t$-design  } & 45 & 42 & 120 & 14190 & 43 & 0.01163 & 0.00167 \\
\hline

\textit{$t$-design  } & 1225 & 1204 & 2800 & $1.17 \times 10^{45}$ & 1223 & 0.0006 & $1.42616 \times 10^{-5}$ \\
\hline

\textit{Transversal Design } & 25 & 20 & 25 & 53130 & 25 & 0.08 & 0.01 \\
\hline

\textit{Transversal Design } & 25 & 20 & 25 & 53130 & 23 & 0.087 & 0.011 \\
\hline

\textit{Transversal Design } & 49 & 42 & 49 & $8.6 \times 10^{7}$ & 47 & 0.04255 &  0.00354\\
\hline

\textit{Subspace Design } &  13 & 10 & 52 & 286 & 13 & 0.115 & 0.023 \\
\hline

\textit{Subspace Design } &  13 & 10 & 52 & 286 & 11 & 0.136 & 0.028 \\
\hline

\textit{Subspace Design } &  130 & 121 & 520 & $2.2 \times 10^{13}$ & 130 & 0.0106 & 0.00057 \\
\hline

\textit{Subspace Design } & 130 & 121 & 520 & $2.2 \times 10^{13}$ & 128 & 0.0108 & 0.00058 \\
\hline

\textit{Subspace Design } & 1210 & 1183 & 4840 & $1.18 \times 10^{55}$ & 1210 & 0.001115 & $1.887 \times 10^{-5}$ \\
\hline

\textit{Subspace Design } & 1210 & 1183 & 4840 & $1.18 \times 10^{55}$ & 1208 & 0.001117 & $1.889 \times 10^{-5}$ \\
\hline

\end{tabular}
\caption{Numerical comparisons between the scheme from \cite{QYS} and designs-based computing schemes presented in this work. The load expression for the scheme in \cite{QYS} is given in Remark \ref{rem:fullstraggleroptimalschemecomparison}, while those of our schemes are compiled in Table \ref{DC_table} (in conjunction with specific constructions of Section \ref{sec:summary_codedcaching}).}
\label{comparison_qys}
\hrule
\end{table*}
\clearpage
\bibliography{IEEEabrv,main.bib}
\bibliographystyle{ieeetr}
\end{document}